\documentclass[aps,prb,reprint,showpacs,superscriptaddress, longbibliography]{revtex4-1}
\pdfoutput=1
\usepackage{mathtools}
\usepackage{graphicx}
\usepackage{amsmath}
\usepackage{amssymb}
\usepackage{amsfonts}
\usepackage{color}
\usepackage{hyperref}

\def\vep{\varepsilon}
\newcommand{\Jpsi}{\psi^{a_l \dots a_1}_{n_l \dots n_1}}
\newcommand{\Jinfpsi}{\psi^{a_l\phantom{-} \dots a_1}_{-n_l \dots -n_1}}
\newcommand{\Jinfpsicaption}{\psi^{a_l\hspace*{0.65em}\dots a_1}_{-n_l \dots -n_1}}
\newcommand{\Jpsib}{\psi^{b_{l'} \dots b_1}_{m_{l'} \dots m_1}}
\newcommand{\JpsiOne}{\psi^{a_{k} \dots a_1}_{1\phantom{_k} \dots 1}}
\newcommand{\JCFT}{(J^{a_l}_{-n_l} \dots J^{a_1}_{-n_1})(0) | 0 \rangle}
\newcommand{\PhiString}{\Phi_{\mathbf{s}}(\mathbf{z})}
\newcommand{\Ja}[1]{J^{a_{#1}}_{-1}}
\newcommand{\NyMin}{N_y^{\text{min}}}
\newcommand{\llangle}{\langle \! \langle}
\newcommand{\rrangle}{\rangle \! \rangle}
\newcommand{\thetaoneopt}{0.0275 \times 2 \pi}
\newcommand{\thetatwoopt}{0.06 \times 2 \pi}
\newcommand{\olap}[2]{\Omega\left(#1, #2\right)}
\newcommand{\compol}{\sqrt{\olap{\psi_G}{\psi_0}^2 + \olap{\psi^0_E}{\psi^z_1}^2}}
\newcommand{\Rth}[4]{\theta\left[\begin{matrix} #1\\#2\end{matrix}\right]\left(#3, #4\right)}

\newcommand{\kwds}{fractional quantum Hall effect; Kalmeyer-Laughlin state; edge states}
\newcommand{\ptitle}{Edge states for the Kalmeyer-Laughlin wave function}
\newcommand{\MPQ}{Max-Planck-Institut f\"ur Quantenoptik, Hans-Kopfermann-Str.\ 1, D-85748 Garching, Germany}

\begin{document}
\author{Benedikt Herwerth}
\email{benedikt.herwerth@mpq.mpg.de}
\affiliation{\MPQ}
\author{Germ\'an Sierra}
\affiliation{Instituto de F\'isica Te\'orica, UAM-CSIC, Madrid, Spain}
\author{Hong-Hao Tu}
\affiliation{\MPQ}
\author{J. Ignacio Cirac}
\affiliation{\MPQ}
\author{Anne E. B. Nielsen}
\affiliation{\MPQ}
\affiliation{Department of Physics and Astronomy, Aarhus University,
 Ny Munkegade 120, DK-8000 Aarhus C, Denmark}

\title{\ptitle}

\begin{abstract}
  We study lattice wave functions obtained from the SU(2)$_1$
Wess-Zumino-Witten conformal field theory. Following Moore and Read's
construction, the Kalmeyer-Laughlin fractional quantum Hall state is
defined as a correlation function of primary fields. By an additional
insertion of Kac-Moody currents, we associate a wave function to each
state of the conformal field theory. These wave functions span the
complete Hilbert space of the lattice system.  On the cylinder, we
study global properties of the lattice states analytically and
correlation functions numerically using a Metropolis Monte Carlo method.
By comparing short-range bulk correlations, numerical evidence is
provided that the states with one current operator represent edge
states in the thermodynamic limit.  We show that the edge states
with one Kac-Moody current of lowest order have a good overlap with
low-energy excited states of a local Hamiltonian, for which the
Kalmeyer-Laughlin state approximates the ground state. For some
states, exact parent Hamiltonians are derived on the cylinder. These
Hamiltonians are SU(2) invariant and nonlocal with up to four-body
interactions.
\end{abstract}


\maketitle

\section{Introduction}
The discovery of the fractional quantum Hall (FQH)
effect\cite{Tsui1982} has revealed the physical existence of a new,
strongly correlated state of matter. One surprising property of FQH
phases is the fact that they cannot be classified in terms of
symmetries\cite{Wen1990}. This is a fundamental difference to most
other known states of matter and led to the concept of classifying
states in terms of topological order\cite{Wen1989,
Wen1990b}. Topological phases are not characterized by a local order
parameter but their properties depend on the topology of the
system. For example, the ground state degeneracy of a FQH system was
shown to depend on the genus of the surface on which the state is
defined\cite{Wen1990}.

A central characteristic of topological order is the appearance of
gapless edge states. Such edge modes are known to be present in a FQH
system\cite{Beenakker1990, MacDonald1990} even though the bulk is
gapped.  It was shown that these edge states can be used to
characterize the topological order of a FQH state\cite{Wen1990c,
Wen1990d}. Furthermore, they are of particular physical importance
because they determine the transport properties of the system. The
strongly correlated state formed by the edge excitations of a FQH
system is that of a chiral Luttinger liquid\cite{Wen1990c} and its
theoretical properties are related to Kac-Moody
algebras\cite{Wen1990c,Wen1991}. For continuous systems, edge state
wave functions were constructed as correlation functions of a
conformal field theory (CFT)\cite{Wen1994, Dubail2012a} and in terms
of Jack polynomials\cite{Lee2014}.

The use of $1+1$ dimensional CFT in
describing FQH states was pioneered by Moore and Read
\cite{Moore1991}. In this approach, a trial wave function for the FQH
state is given as the correlation function of conformal primary
fields. Laughlin's wave function for the continuum\cite{Laughlin1983}
as well as its SU(2) symmetric, bosonic lattice version, the
Kalmeyer-Laughlin (KL) state\cite{Kalmeyer1987, Kalmeyer1989}, were
shown to be of this form\cite{Moore1991, Tu2014b, Nielsen2012,
Nielsen2013d}.  It was already conjectured by Moore and
Read\cite{Moore1991} that the FQH edge modes should be described by
the same CFT that defines the bulk wave function.  The idea of this
bulk-edge correspondence was later developed into an extension of
Moore and Read's method\cite{Wen1994, Dubail2012a}. Within this
approach, trial FQH edge states are obtained from CFT descendant
states.

This paper is based on a similar ansatz for edge states in lattice
systems. Our starting point is the SU(2)$_1$ Wess-Zumino-Witten (WZW)
theory, which was used previously\cite{Nielsen2012, Nielsen2013d} to
obtain the KL state as a correlation function of primary fields. By insertion of
Kac-Moody currents, we define a tower of states, which
corresponds to the CFT descendant states. Since we work in a spin
formulation, the Hilbert space of the lattice system is that of $N$
spin-$\frac{1}{2}$ degrees of freedom, where occupied sites are
represented as spin-up and empty sites as spin-down.

We show that the mapping from CFT states to spin states is surjective,
i.e. any state of the spin system can be written as a linear
combination of states constructed from the CFT.  As a consequence, not
all states obtained in this way are edge states.  For some
of the wave functions, we carry out numerical calculations to
test if they describe edge states: On the cylinder, we
compare spin correlation functions in the states with one current
operator to the KL state using a
Metropolis Monte Carlo algorithm\cite{Metropolis1953,
Handscomb1962}.  We show that the nearest-neighbor bulk correlations
approach each other exponentially as the thermodynamic limit is
taken. This indicates that the states with one current operator indeed
describe edge states.

In the past years, parent Hamiltonians of the KL
state\cite{Schroeter2007, Thomale2009, Nielsen2012, Greiter2014} and its non-Abelian
generalizations\cite{Paredes2009, Greiter2009, Nielsen2011, Greiter2014, Tu2013c,
  Glasser2015} were constructed. It was also shown that the KL state\cite{Nielsen2013a}
and non-Abelian FQH lattice states\cite{Greiter2009, Glasser2015} have
a good overlap with the ground states of certain local Hamiltonians.
For one of these Hamiltonians, an implementation scheme in ultracold atoms in optical lattices was
proposed\cite{Nielsen2013a}. In this work, we study the local Hamiltonians of
Ref.~\onlinecite{Nielsen2013a} on the cylinder and use the KL state and
the edge states with one current operator of order one as ansatz
states. We find that there is a choice of parameters for which both
the ground state and some low-energy excited states have a good
overlap with our ansatz states.

Furthermore, we construct Hamiltonians for which the KL state and some of the
states obtained from it by insertion of current operators are exact
ground states.  The proposed parent Hamiltonians are SU(2)
invariant and nonlocal with up to four-body interactions.

This paper is structured as follows: In Sec.~\ref{sec:CFT-model}, we
introduce the CFT model, define the map to a spin-$\frac{1}{2}$
system, and show that it is surjective. In
Sec.~\ref{sec:properties-on-the-cylinder}, properties of our states on
the cylinder are studied and numerical evidence is given that the
states with one current operator describe edge states. We consider a
local model Hamiltonian in Sec.~\ref{sec:local-model-hamiltonians}
and derive exact parent Hamiltonians for some states on
the cylinder in Appendix~\ref{sec:exact-parent-hamiltonians}.
The conclusion is given in Sec.~\ref{sec:conclusion}.

\section{CFT model and spin-$\frac{1}{2}$ states}
\label{sec:CFT-model}
In this section, we review some properties of the
SU(2)$_1$ Wess-Zumino-Witten (WZW) theory and
define the correspondence between states of the
CFT and states of a spin-$\frac{1}{2}$ system on the lattice.
It is shown that this map of CFT states to lattice
spin states is surjective.

\subsection{SU(2)$_1$ Wess-Zumino-Witten theory}
We consider the chiral sector of the SU(2)$_1$ WZW
theory. In addition to conformal invariance, this theory exhibits an
SU(2) symmetry generated by the currents $J^a(z)$ with $a \in \{x, y, z\}$.
The modes $J^a_n$ are defined in terms of the
Laurent expansion
\begin{align}
\label{eq:laurent-expansion-of-current-operators}
  J^a(z) &= \sum_{n=-\infty}^{\infty} J^a_n z^{-n-1},\notag\\
  J^a_n &= \oint_0 \frac{\mathrm{d} z}{2 \pi i} z^{n} J^a(z),
\end{align}
and satisfy the
Kac-Moody algebra \cite{Knizhnik1984}
\begin{align}
\label{eq:Kac-Moody-current-algebra} \left[J^a_m, J^b_n\right] &= i
\vep_{abc} J^c_{m+n} + \frac{m}{2} \delta_{ab} \delta_{m+n, 0}.
\end{align}
Here, $\vep_{abc}$ is the Levi-Civita symbol and $\delta_{ab}$ the
Kronecker delta. For indices $c \in \{x, y, z\}$, we adopt the
convention that indices occurring twice are summed over, unless
explicitly stated otherwise.

The SU(2)$_1$ WZW theory has two primary fields, the identity with
conformal weight $h=0$, and a two-component field $\phi_s(z)$ ($s=\pm
1$) with conformal weight $h=\frac{1}{4}$.  The field $\phi_s(z)$
provides a spin-$\frac{1}{2}$ irreducible representation through its
operator product expansion (OPE) with the SU(2)
currents\cite{DiFrancesco1997}:
\begin{align}
\label{eq:ope-J-primaries}
J^a(z) \phi_s(w) &\sim -\frac{1}{z-w} \sum_{s'=\pm 1} t^a_{s s'} \phi_{s'}(w),
\end{align}
where $t^a = \frac{\sigma^a}{2}$ are the SU(2) spin operators.

The field content of the SU(2)$_1$ WZW theory can be represented in terms
of the chiral part $\varphi(z)$ of a free boson field as
\begin{align}
\label{eq:free-boson-representation}
\phi_s(z) &= e^{i \pi (q-1) (s+1) / 2} :e^{i s \varphi(z)/ \sqrt{2}}:, \notag \\
J^z(z) &= -\frac{i}{\sqrt{2}} \partial_z \phi(z), \notag \\
J^{\pm}(z) &= J^x(z) \pm i J^y(z) = e^{i \pi (q-1)} :e^{\mp i \sqrt{2} \varphi(z)}:.
\end{align}
Here, $q=0$ if the operators act on a state with an even number of
modes of the $h=\frac{1}{4}$ primary field and $q=1$ otherwise, and
the colons denote normal ordering. The value of $s \in \{-1, 1\}$ equals two
times the spin-$z$ eigenvalue of $\phi_s(z)$.

A general state in the identity sector of the CFT Hilbert space is a linear combination
of states
\begin{align}
  \label{eq:cft-state}
  \JCFT,
\end{align}
where $| 0 \rangle$ is the CFT vacuum and $n_i$ are positive integers.
The sum $k =
  \sum_{i=1}^{l} n_i$ defines the level of the state. By means of the
Kac-Moody algebra of Eq.~\eqref{eq:Kac-Moody-current-algebra}, a
basis can be chosen for which the mode numbers are ordered: $n_l \ge
  n_{l-1} \ge \dots \ge n_1 > 0$.

\subsection{Spin states on the lattice}
\label{sec:lattice-spin-one-half-states}
To each CFT state $\JCFT$, we associate a state
\begin{align}
\label{eq:state-in-spin-one-half-Hilbert-space}
|\Jpsi \rangle &= \sum_{s_1, \dots, s_N} \Jpsi(s_1, \dots, s_N) |s_1, \dots, s_N\rangle
\end{align}
in the Hilbert space of a system of $N$ spin-$\frac{1}{2}$ degrees of freedom.
Its spin wave function is defined as the CFT correlator
\begin{align}
\label{eq:spin-wave-function-as-correlator}
  &\Jpsi(s_1, \dots, s_N) \notag \\
  &\quad= \langle \phi_{s_1}(z_1) \dots \phi_{s_N}(z_N) (J^{a_l}_{-n_l} \dots J^{a_1}_{-n_1})(0) \rangle,
\end{align}
where $\langle \dots \rangle$ denotes the expectation
value of radially ordered operators in the CFT vacuum.

In the sum of Eq.~\eqref{eq:state-in-spin-one-half-Hilbert-space},
$s_i = \pm 1$ and $|s_1, \dots, s_N \rangle$ is the tensor product of
eigenstates $|s_i\rangle$ of the spin operator $t^z_i$ at position $i$
($t^z_i | s_i \rangle = \frac{s_i}{2} | s_i \rangle$).  The complex
coordinates $z_i$ are parameters of the wave function $\Jpsi(s_1,
\dots, s_N)$ and define a lattice of positions in the complex plane.
Since we want to keep them fixed, we do not explicitly indicate the
parametric dependence of $\Jpsi(s_1, \dots, s_N)$ on the positions
$z_i$ for simplicity of notation.

The wave function corresponding to the CFT vacuum is given by
\cite{DiFrancesco1997}
\begin{align}
\label{eq:vertex-op-correlator}
\psi_0(s_1, \dots, s_N) &\equiv \langle \phi_{s_1}(z_1) \dots \phi_{s_N}(z_n) \rangle \\
&= \delta_{\mathbf{s}} \chi_{\mathbf{s}} \prod_{i < j}^{N} (z_i - z_j)^{s_i s_j / 2}\notag,
\end{align}
where $\delta_{\mathbf{s}}$ is $1$ if $\sum_{i=1}^N s_i = 0$ and $0$
otherwise. $\chi_{\mathbf{s}}$ denotes the Marshall sign factor,
\begin{align}
\label{eq:marshall-sign-factor}
\chi_{\mathbf{s}} &= \prod_{q=1}^{N} e^{i \pi (q-1) (s_q+1)/2}.
\end{align}
It follows from the alternating sign in the definition of $\phi_s(z)$
[cf. Eq.~\eqref{eq:free-boson-representation}]. Due to the presence
of the Marshall sign factor, the state $\psi_0$ is a singlet
of the total spin\cite{Nielsen2011} $T^a = \sum_{i=1}^N t^a_i$:
\begin{align}
  \label{eq:psi0-singlet-property}
  T^a | \psi_0 \rangle = 0.
\end{align}
We assume that $N$ is even, as required by the charge neutrality
condition $\sum_{i=1}^N s_i = 0$ in the wave function of $\psi_0$.
It was shown previously that $\psi_0$ is equivalent to the
KL state\cite{Nielsen2012, Tu2014b}.
We note that the wave function $\psi_0$ of Eq.~\eqref{eq:vertex-op-correlator}
contains a square root, which has two branches. It is assumed that the same
branch is taken for all factors in Eq.~\eqref{eq:vertex-op-correlator}. Which
of the two branches is chosen is not important, since this choice only influences
the wave function by a total factor.

The states $\Jpsi$ can be related to $\psi_0$ by an application of
spin operators \cite{Herwerth2015}: In
Eq.~\eqref{eq:spin-wave-function-as-correlator}, one can write the
modes $J^{a_i}_{-n_i}$ as an integral according to
Eq.~\eqref{eq:laurent-expansion-of-current-operators}, deform the
integral contour, and apply the OPE~\eqref{eq:ope-J-primaries} between
current operators and primary fields.
Defining
\begin{align}
\label{eq:definition-of-u-operators}
u^a_{k} & \equiv \sum_{i=1}^N (z_i)^{k} t^a_i
\end{align}
it then follows that
\begin{align}
\label{eq:jpsi-from-psi0}
| \Jpsi \rangle &= u^{a_l}_{-n_l} \dots u^{a_1}_{-n_1} | \psi_0 \rangle.
\end{align}

We note that the operators $u^a_k$ defined in
Eq.~\eqref{eq:definition-of-u-operators} satisfy
\begin{align}
  \label{eq:commutator-algebra-of-u-operators}
  \left[u^a_m, u^b_n\right] &= i \vep_{abc} u^c_{m + n},
\end{align}
which is a Kac-Moody algebra with vanishing central
extension.

In addition to the states $\Jpsi$, we also consider the wave functions obtained
by the insertion of two additional primary fields, one at $z_0=0$ and one
at $z_\infty=\infty$:
\begin{align}
\label{eq:psi0-0inf}
  &\psi_0^{s_0, s_\infty}(s_1, \dots, s_N) \notag \\
  &\quad \equiv \langle \phi_{s_\infty}(\infty) \phi_{s_1}(z_1) \dots \phi_{s_N}(z_N) \phi_{s_0}(0) \rangle \notag \\
  &\quad \propto \delta_{\bar{\mathbf{s}}} \chi_{\mathbf{s}} (-1)^{(1-s_\infty)/2} \prod_{n=1}^{N} z_n^{s_0 s_n / 2} \prod_{n < m}^N (z_n - z_m)^{s_n s_m / 2},
\end{align}
where $\delta_{\bar{\mathbf{s}}}$ is $1$ if $s_0 + s_{\infty} + \sum_{j=1}^N s_j = 0$
and $0$ otherwise. The state $\psi_0^{s_0, s_\infty}$ contains a singlet and a triplet
obtained by the tensor product decomposition of the two
additional spin-$\frac{1}{2}$ fields. As we show in
Sec.~\ref{sec:relation-to-torus}, the singlet component on the
cylinder can be derived from the wave function of $N$ primary fields
on a torus in the limit where the torus becomes a cylinder.

We note that the states $\psi_0$, $\Jpsi$, and $\psi^{s_0,
s_\infty}_0$ of
Eqs.~(\ref{eq:spin-wave-function-as-correlator},~\ref{eq:vertex-op-correlator}),
and~\eqref{eq:psi0-0inf} are not normalized. Whenever the norm of a
state is needed, it will be explicitly included in the corresponding
expression.

\subsection{Completeness of spin states}
\label{sec:completeness-of-spin-states}
One may ask if the linear transformation that
maps CFT states $\JCFT$ to
spin states $\Jpsi$ is surjective, i.e. whether
any state in the Hilbert space $\mathcal{H}_N$ of $N$ spin-$\frac{1}{2}$ particles
can be written as a linear combination of the states $\Jpsi$.
We now show that this is indeed the case. Introducing the $N \times N$
matrix
\begin{align}
\label{eq:mat-spin-operator-basis-transformation}
\mathcal{Z} &= \left(
\begin{matrix}
1 &\dots &1\\
(z_1)^{-1} &\dots &(z_N)^{-1}\\
(z_1)^{-2} &\dots &(z_N)^{-2}\\
\vdots& \vdots & \vdots \\
(z_1)^{-(N-1)} &\dots &(z_N)^{-(N-1)}
\end{matrix}
\right),
\end{align}
the definition of Eq.~\eqref{eq:definition-of-u-operators} becomes
\begin{align}
  \label{eq:definition-of-u-operators-in-matrix-form}
  \left(\begin{matrix} u^a_0\\u^a_{-1}\\ \vdots\\u^a_{-(N-1)} \end{matrix}\right) =
  \mathcal{Z}
  \left(\begin{matrix} t^a_1\\t^a_2\\ \vdots\\t^a_N \end{matrix}\right).
\end{align}
The determinant of $\mathcal{Z}$ is the well known Vandermonde determinant:
\begin{align}
  \label{eq:vandermonde-det}
  \mathrm{det}\left(\mathcal{Z}\right) = \prod_{i < j}^N (z_j^{-1} - z_i^{-1}).
\end{align}
$\mathcal{Z}$ is therefore nonsingular if all positions $z_i$ are distinct, which
we assume to be the case. As a consequence, the relation
of Eq.~\eqref{eq:definition-of-u-operators-in-matrix-form}
can be inverted, i.e. all spin operators $t^a_j$ can be written
as linear combinations of $u^a_{-n}$ with $n \in \{0, \dots, N-1\}$.

The Hilbert space $\mathcal{H}_N$ is spanned by the states obtained
from any nonzero state by successive application of spin operators $t^a_i$.
Given that $t^a_i$ can be expressed in terms of $u^a_{-n}$, it follows in particular
that the states $| \Jpsi \rangle = u^{a_l}_{-n_l} \dots u^{a_1}_{-n_1} | \psi_0 \rangle$ with $n_i \in \{0, \dots, N-1\}$
span $\mathcal{H}_N$. Since $\psi_0$ is a singlet and $u^a_0 = \sum_{j=1}^N t^a_j$ is the total spin,
a state
\begin{align}
  \label{eq:commute-u0-to-the-right}
  u^a_0 u^{a_l}_{-n_l} \dots u^{a_1}_{-n_1} | \psi_0 \rangle
\end{align}
can be written in terms of states for which all mode numbers
are greater than zero by commuting $u^a_0$ to the right until
it annihilates $\psi_0$ [cf. the commutator of
Eq.~\eqref{eq:commutator-algebra-of-u-operators}].
Therefore, $\mathcal{H}_N$ is spanned
by the states $\Jpsi$ with $n_i > 0$.

This argument shows that not all states constructed by the insertion of current
operators can be edge states compared to $\psi_0$.
For the states with one current operator $J^a_{-n}$, numerical
evidence will be given in Sec.~\ref{sec:spin-correlation-functions} that
these represent edge states on the cylinder.

The fact that the states $\Jpsi$ span $\mathcal{H}_N$ raises the
question about the minimal level $k =\sum_{j=1}^l n_j$ needed to
obtain the complete Hilbert space. We note that an upper bound is
given by $k = N (N-1)$, since any product of spin operators $t^a_i$
can be reduced to a product of at most $N$ spin operators.
Each of these spin operators can then be expanded in
terms of the operators $u^a_{-n}$ with $n \in \{0, \dots, N-1\}$.  We
carried out numerical calculations for the states $\Jpsi$ which indicate
that the states up to level $k=(N/2)^2$ are enough to obtain the
complete Hilbert space from $\psi_0$.

\section{Properties of states on the cylinder}
\label{sec:properties-on-the-cylinder}
In this section, we study properties of the states defined in
Sec.~\ref{sec:lattice-spin-one-half-states} on the cylinder.

We consider a square lattice with $N_x$ lattice sites in the open
direction and $N_y$ lattice sites in the periodical direction
of the cylinder.
After mapping the cylinder to the complex plane,
the coordinates assume the form
\begin{align}
  \label{eq:lattice-positions-on-complex-plane}
  z_{j} = e^{\frac{2 \pi}{N_y} (j_x + i j _y)} e^{-\frac{2 \pi}{N_y} \frac{N_x + 1}{2}}.
\end{align}
Here, $j_x \in \{1,\dots, N_x\}$ is the $x$-component of the index and
$j_y \in \{1, \dots, N_y\}$ is the $y$-component, so that
$j = (j_x - 1) N_y + j_y$ ranges from $1$ to $N = N_x N_y$.
For the remainder of this paper, we adopt a two-index notation, where this
is convenient, i.e. we may write $z_{j_x, j_y}$ instead of $z_j$, denoting
the $x$- and $y$-components by subscripts.

Note that one has the freedom to rescale the coordinates since this
changes the wave functions only by a total factor.  The constant
factor that we included in
Eq.~\eqref{eq:lattice-positions-on-complex-plane} is chosen such that
the center of the cylinder is at the unit circle.

We assume that the number of sites $N$ is even.  It is also possible
to study the case of $N$ being odd which will show the existence of
two topological sectors.  However, we can already
identify the two anyonic sectors for $N$ even.  As will be shown below in
Sec.~\ref{sec:relation-to-torus}, the state $\psi_0$ and the singlet
component $\psi^{\text{sgl}}_0$ of the state with two additional spins
(one at $z_0 = 0$ and one at $z_\infty=\infty$) can be obtained from
the wave function of $N$ primary fields on the torus in the limit
where the torus becomes a cylinder.  This argumentation shows that the
two states $\psi_0$ and $\psi^{\text{sgl}}_0$ represent the two
anyonic sectors in the case of an even number of spins. It would be
possible to consider an odd number of sites on the cylinder by putting
an additional spin either at $z_0 = 0$ or at $z_\infty=\infty$ so that
the charge neutrality condition is satisfied. On the torus, however,
such a construction is not possible and the argumentation that we used
to identify the two sectors for even $N$ does not directly apply.

\subsection{Global transformation properties of CFT states}
\label{sec:global-properties}
In this subsection, we study global transformation properties of the
states $\psi_0$, $\Jpsi$, and $\psi_0^{s_0, s_\infty}$. This serves
two purposes: First, it allows us to conclude that states with a
different momentum are orthogonal, i.e. they have different global
properties. Later, we will study their local behavior numerically and
compare spin correlation functions in the bulk. The symmetries derived
in this subsection will be exploited in our numerical calculations
to obtain efficient Monte Carlo estimates.

We consider the translation operator in the periodical direction
$\mathcal{T}_y$ and the inversion operator $\mathcal{I}$. Their
precise definition and the derivation of their action on the
states $\psi_0$, $\Jpsi$, and $\psi_0^{s_0, s_\infty}$ are given in
Appendix~\ref{sec:appendix-derivation-of-global-properties}.
Geometrically, the translation operator rotates the system in the
periodical direction and the inversion operator corresponds to a
reflection of the cylinder along its two central cross sections. We
call it an inversion because it acts on the coordinates defined in
Eq.~\eqref{eq:lattice-positions-on-complex-plane} as $z_i \to
z_i^{-1}$.

Eigenstates of $\mathcal{T}_y$ and $\mathcal{I}$ are given in
Table~\ref{tab:global-transformation-properties}.  As we show in
Appendix~\ref{sec:appendix-inversion-operator}, applying the inversion
$\mathcal{I}$ to $\Jpsi$ corresponds to inserting the current operators
at $z_{\infty}=\infty$ instead of $z_0=0$. We use the notation $\Jinfpsi$
for these states:
\begin{align}
  \label{eq:definition-of-Jpsi-inf}
  &\Jinfpsi(s_1, \dots, s_N) \notag\\
  &\quad\equiv \langle J^{a_1}_{n_1} \dots J^{a_l}_{n_l} \phi_{s_1}(z_1)\dots \phi_{s_N}(z_1) \rangle \notag\\
  &\quad= (u^{a_l}_{n_l} \dots u^{a_1}_{n_1} \psi_0)(s_1, \dots, s_N).
\end{align}
Since the momentum in the periodical direction $P_y$ is related to
$\mathcal{T}_y$ through the relation $\mathcal{T}_y = e^{i P_y}$, we
conclude from Table~\ref{tab:global-transformation-properties} that an
additional insertion of a current operator $J^a_{-n}$ into the
correlation function of primary fields adds a momentum of $-2 \pi
n/N_y$ to the state. In particular, the states $\psi_0$ and $\Jpsi$
have a different momentum if $k = \sum_{j=1}^l n_j$ is different from
$0$ modulo $N_y$.

\begin{table}[htb]
  \caption{\label{tab:global-transformation-properties}
    Eigenstates of the translation operator $\mathcal{T}_y$ and the
inversion $\mathcal{I}$. The sum of mode numbers $\sum_{j=1}^l n_j$ is
denoted by $k$.  For the states $\Jinfpsicaption$, the current operators are
inserted at $z_\infty = \infty$ [cf. Eq.~\eqref{eq:definition-of-Jpsi-inf}].}
\centering
\begin{ruledtabular}

\begin{tabular}{lll}
                                                                                           Eigenstate &                                     $\mathcal{T}_y$ &                      $\mathcal{I}$ \\
\hline
                                                                                             $\psi_0$ &                            $(-1)^{N_x \frac{N}{2}}$ &           $(-1)^{N_y \frac{N}{2}}$ \\
                                                                                              $\Jpsi$ & $(-1)^{N_x \frac{N}{2}} e^{-\frac{2 \pi i}{N_y} k}$ &                                --- \\
                                                                                 $\Jpsi \pm \Jinfpsi$ &                                                 --- &    $(\pm 1)(-1)^{N_y \frac{N}{2}}$ \\
                                                                            $\psi^{s_0, s_\infty}_0$: &                      $(-1)^{N_x \frac{N}{2} + N_x}$ &                                --- \\
                                 $\quad\psi^{\uparrow, \downarrow}_0 - \psi^{\downarrow, \uparrow}_0$ &                      $(-1)^{N_x \frac{N}{2} + N_x}$ &     $(-1)^{N_y \frac{N}{2} + N_x}$ \\
 $\quad \psi^{\uparrow, \uparrow}_0,  \psi^{\uparrow, \downarrow}_0 + \psi^{\downarrow, \uparrow}_0,$ &                     $(-1)^{N_x \frac{N}{2} + N_x }$ & $(-1)^{N_y \frac{N}{2} + N_x + 1}$ \\
                                                 $\quad \text{ and } \psi^{\downarrow, \downarrow}_0$ &                                                     &                                    
\end{tabular}
  \end{ruledtabular}
\end{table}

\subsection{Relation to the KL states on the torus}
\label{sec:relation-to-torus}
In this subsection, we place the system on the torus and take a limit in which
the torus becomes a cylinder.  We show that the wave function of $N$ primary
fields on the torus gives rise to $\psi_0$ and the singlet component $\psi^{\mathrm{sgl}}_0$
of $\psi^{s_0, s_\infty}_0$ on the cylinder.

\begin{figure}[htb]
\centering
\includegraphics{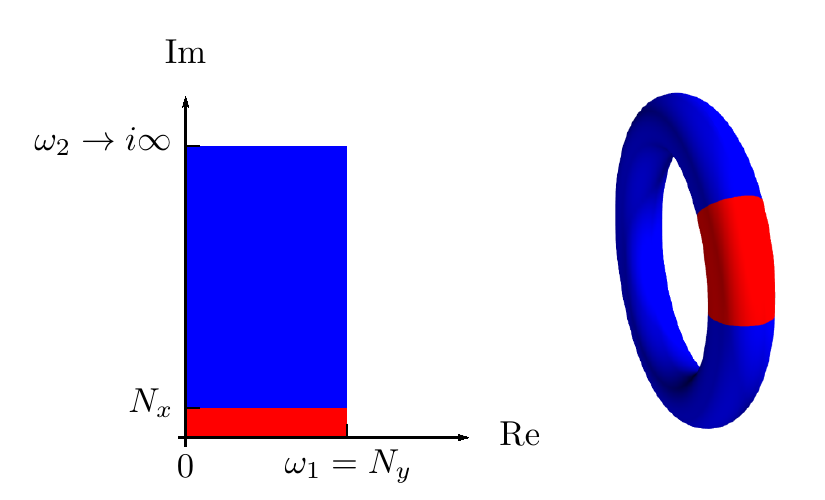}
\caption{\label{fig:torus-to-cylinder}
  (Color online) Limit in which the torus becomes a cylinder:
  The circumference $|\omega_2|$ is taken to infinity while the positions
  lie in the finite region of size $N_x \times N_y$ (red patch).}
\end{figure}

We define the torus for $\omega_1 > 0$ and $\omega_2 = i L$ with $L >
0$ by identifying a complex number $z$ with $z + n \omega_1 + m
\omega_2$ for $m, n \in \mathbb{Z}$. The two circumferences of the torus are
therefore given by $\omega_1$ and $|\omega_2|$.  Let us denote the
positions on the torus by $v_i$, i.e. we assume that $v_i$ lie in the rectangle
spanned by $\omega_1$ and $\omega_2$.
Keeping the positions $v_i$ fixed and taking the circumference $L = |\omega_2| \to \infty$
transforms the torus into a cylinder, as illustrated in Fig.~\ref{fig:torus-to-cylinder}.

On the torus, there are two states $\psi^{\mathrm{torus}}_k$ with $k \in \{0, \frac{1}{2}\}$.
These are given by\cite{Nielsen2013d}
\begin{align}
  \label{eq:psi0-on-torus}
  &\psi^{\mathrm{torus}}_k(s_1, \dots, s_N) = \langle \phi_{s_1}(v_1) \dots \phi_{s_N}(v_N) \rangle_k \notag \\
  &\propto \delta_{\mathbf{s}} \chi_{\mathbf{s}} \Rth{k}{0}{\sum_{i=1}^{N} \zeta_i s_i}{2 \tau}
   \prod_{i < j}^N \left(\Rth{\frac{1}{2}}{\frac{1}{2}}{\zeta_i - \zeta_j}{\tau}\right)^{s_i s_j/2}.
\end{align}
Here, $\zeta_i = v_i/\omega_1$ are the rescaled positions, $\tau =
\omega_2/\omega_1$ is the modular parameter of the torus,
and $\theta$ the Riemann theta function defined
as
\begin{align}
  \label{eq:theta-function}
  \Rth{a}{b}{\zeta}{\tau} &= \sum_{n \in \mathbb{Z}} e^{i \pi \tau (n + a)^2 + 2 \pi i (n + a) (\zeta + b)}.
\end{align}
The limit $\omega_2 \to i \infty$, which transforms the torus into a cylinder,
implies $\tau \to i \infty$. In this case, only the
terms with the smallest value of $(n+a)^2$ contribute to the sum of Eq.~\eqref{eq:theta-function}.
These terms have $n=0$ for $a=0$ and $n \in \{-1, 0\}$ for $a=\frac{1}{2}$. Therefore,
\begin{alignat}{3}
  \label{eq:torus-to-cylinder-limit-of-terms}
  &\Rth{0}{0}{\zeta}{2 \tau} &\to& 1, \notag \\
  &\Rth{\frac{1}{2}}{0}{\zeta}{2 \tau} &\to& e^{\frac{\pi i \tau}{2}} (e^{i \pi \zeta} + e^{-i \pi \zeta}), \quad \text{and}\notag \\
  &\Rth{\frac{1}{2}}{\frac{1}{2}}{\zeta_i - \zeta_j}{\tau} &\to& i e^{-i \pi (\zeta_i + \zeta_j)} e^{\frac{i \pi \tau}{4}} (e^{2 \pi i \zeta_i} - e^{2 \pi i \zeta_j}),
\end{alignat}
for $\tau \to i \infty$.
With $\sum_{j=1}^N s_j = 0$, it follows that
\begin{align}
  \label{eq:product-over-plus-term}
  \prod_{m < n}^N e^{-i \frac{\pi}{2} (\zeta_m + \zeta_n) s_m s_n} &= e^{i \frac{\pi}{2} \sum_{n=1}^N \zeta_n}.
\end{align}
In the limit $\omega_2 \to i \infty$, we therefore obtain
\begin{align}
  \label{eq:limit-torus-cylinder-psi0}
  &\psi^{\mathrm{torus}}_0(s_1, \dots, s_N) \notag \\
  &\propto \delta_{\mathbf{s}} \chi_{\mathbf{s}} \prod_{m < n}^N \left(e^{2 \pi i v_m/\omega_1} - e^{2 \pi i v_n/\omega_1}\right)^{s_m s_n / 2}
\end{align}
and
\begin{align}
  \label{eq:limit-torus-cylinder-psi0inf-sgl}
  &\psi^{\mathrm{torus}}_{\frac{1}{2}}(s_1, \dots, s_N) \notag \\
  &\propto \delta_{\mathbf{s}} \chi_{\mathbf{s}} \left(\prod_{n=1}^{N} e^{\pi i v_n s_n / \omega_1} + \prod_{n=1}^{N} e^{-\pi i v_n s_n / \omega_1}\right)\notag \\
  &\phantom{\underset{\omega_2 \to i \infty}{\propto}} \times \prod_{m < n}^N \left(e^{2 \pi i v_m/\omega_1} - e^{2 \pi i v_n/\omega_1}\right)^{s_m s_n / 2}.
\end{align}
The exponentials $e^{2 \pi i v_n / \omega_1}$ lie on a
cylinder of circumference $\omega_1$. We therefore identify $z_n = e^{2 \pi i
  v_n / \omega_1}$ and $N_y = \omega_1$. Comparing the expressions
for $\psi^{\text{torus}}_k$ in the limit $\omega_2 \to \infty$ to
Eqs. \eqref{eq:vertex-op-correlator} and \eqref{eq:psi0-0inf}, we
conclude that $\psi^{\mathrm{torus}}_0 \propto \psi_0$ and
$\psi^{\mathrm{torus}}_{\frac{1}{2}} \propto \psi^{\uparrow,
\downarrow}_0 - \psi^{\downarrow, \uparrow}_0 \equiv
\psi^{\mathrm{sgl}}_0$.

It would be very interesting to investigate the relation between excited states in one
and two dimensions on the circle and the plane, respectively, to those on the torus.
The result for the relation between $\psi^{\text{torus}}_{\frac{1}{2}}$ and $\psi^{\text{sgl}}_0$
is a first step in that direction.

\subsection{Spin correlation functions and edge states}
\label{sec:spin-correlation-functions}
We calculated two-point spin correlation functions in the states $\psi_0$, $\psi^a_1$
and in the singlet state
\begin{align}
\label{eq:definition-of-psi0-sgl}
\psi^{\text{sgl}}_0 \equiv \psi^{\uparrow, \downarrow}_0 - \psi^{\downarrow, \uparrow}_0
\end{align}
using a Metropolis Monte Carlo algorithm. This allowed us to compare
properties of the states numerically for large system sizes by sampling
the relevant probability distributions. We furthermore exploited the
translation and inversion symmetries of Table~\ref{tab:global-transformation-properties}
to average over equivalent correlation functions, thus obtaining a faster
converging Monte Carlo estimate.

\begin{figure*}[htb]
\centering
\includegraphics{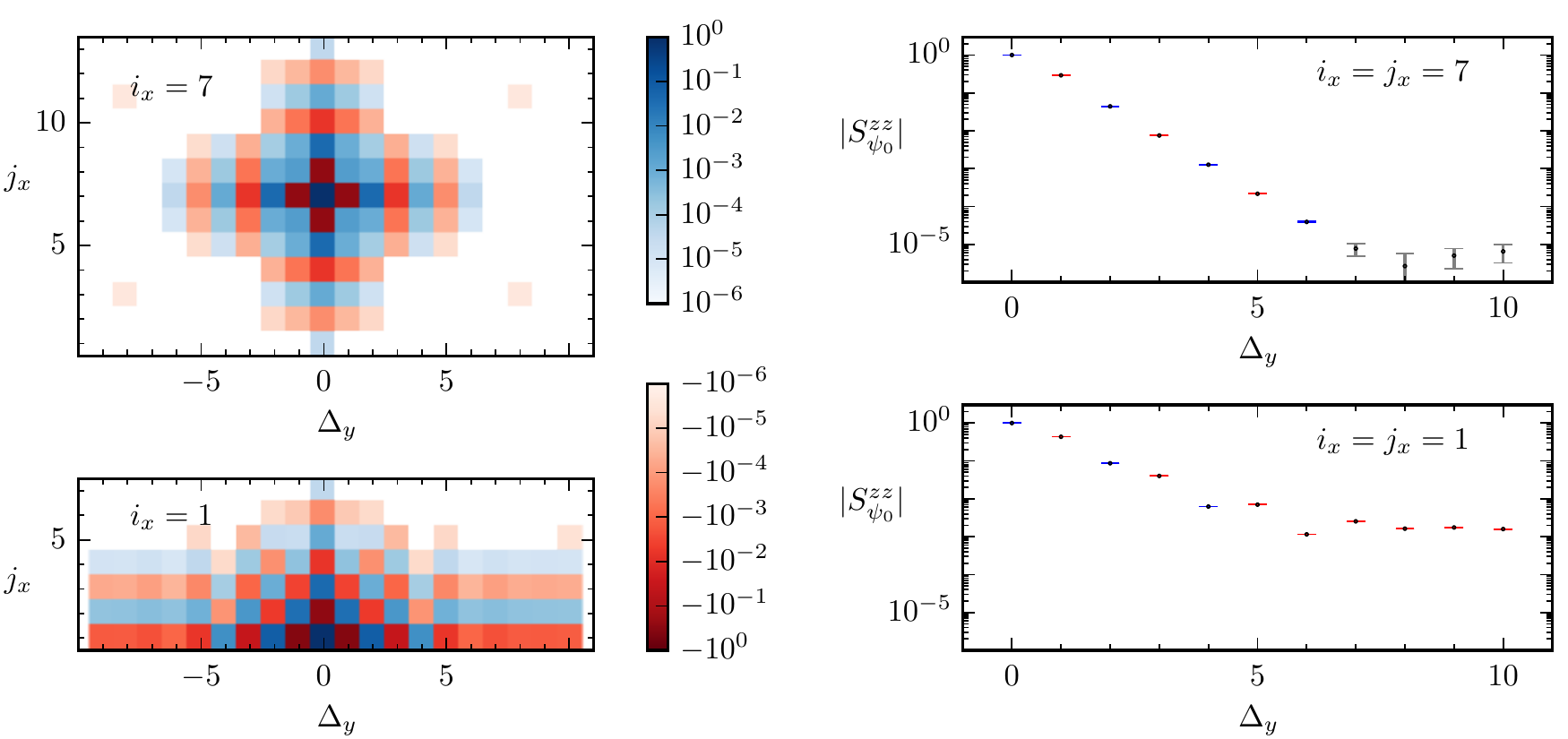}
\caption{\label{fig:psi0-2d-zz-correlations} (Color online) Two-point
spin correlation function $S^{zz}_{\psi_0}(i_x, j_x, \Delta_y)$ in the
bulk (upper panels) and at the edge (lower panels) for $N_x = 13$ and
$N_y = 20$. The left panels show the two-dimensional dependency in a
color plot. Whenever the value of the correlation function does not
differ from zero by more than three times the estimated error, we
excluded the data point from the plot (blank fields). In the right
panels, the absolute value $|S^{zz}_{\psi_0}(i_x, i_x, \Delta_y)|$ of
the correlation function along the $y$-direction is plotted.  Points
for which the sign of the correlation function is positive (negative)
are shown in blue (red). For the data shown in gray, the mean value
does not differ from zero by more that three times the estimated
error. In the bulk, the correlations decay exponentially, while a
nonzero, negative correlation remains at the edge for $\Delta_y \ge 5
$.}
\end{figure*}

In this subsection, we shall use the notation
\begin{align}
\label{eq:tow-point-function}
  S^{ab}_\psi(i_x, j_x, \Delta_y)
  = 4 \frac{\langle \psi | t^a_{i_x, \Delta_y + 1} t^b_{j_x, 1} | \psi \rangle}{\langle \psi | \psi \rangle}
\end{align}
for the two-point correlation function in a state $\psi$. Since all
wave functions that we consider have a translational symmetry in the
periodical direction, their value only
depends on the difference $\Delta_y$ of the positions in the
$y$-direction.

Before comparing the wave functions with each other, we discuss the
spin ordering pattern in $\psi_0$, which is encoded in the correlation
function $S^{zz}_{\psi_0}(i_x, j_x, \Delta_y)$. (Since $\psi_0$ is a
singlet, $xx$-, $yy$- and $zz$-correlations are the same and only
correlation functions with $a=b$ are nonzero.)  Our numerical results
are shown in Fig.~\ref{fig:psi0-2d-zz-correlations}.  In the bulk of
the system, we observe a ring-like structure with an alternating
magnetization. At the edge, the correlations are anti-ferromagnetic at
short distances.  At larger distances along the $y$-direction,
however, the sign becomes stationary and a negative correlation remains.
In the two-dimensional picture, the ordering is still characterized
by an alternating magnetization with the sign of the correlation
function changing along the $x$-direction.

\begin{figure}[htb]
\centering
\includegraphics[width=1.0\linewidth]{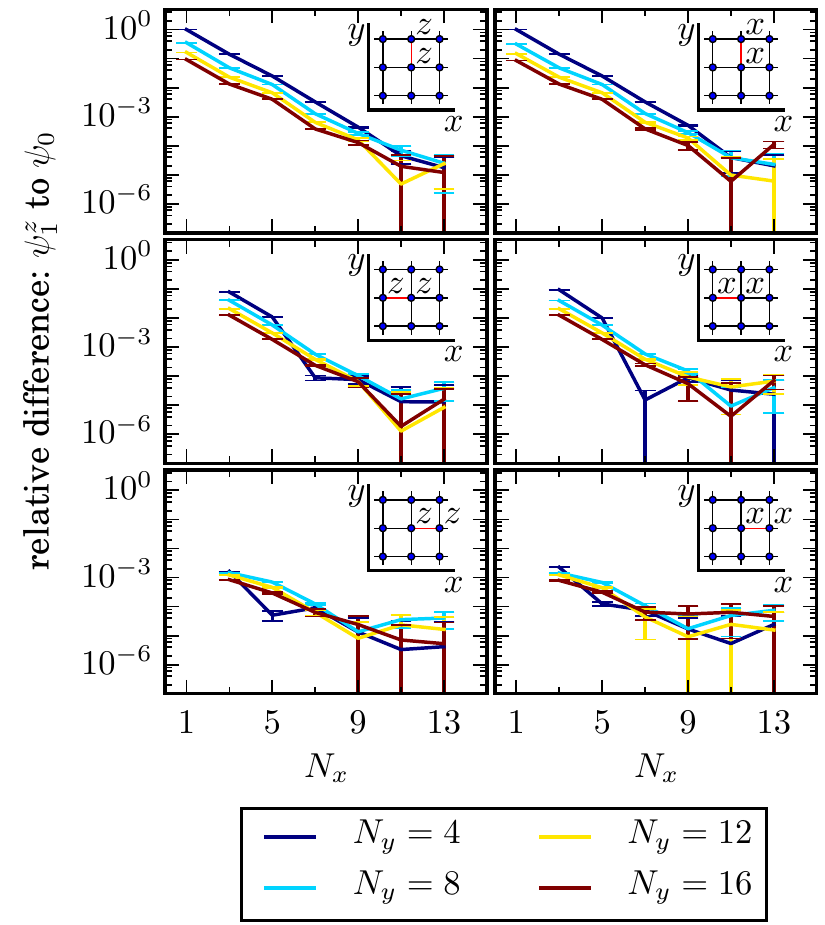}
\caption{\label{fig:psi0-vs-uz-psi0} (Color online)
  Comparison of nearest-neighbor bulk correlations in
  $\psi^z_1$ and $\psi_0$. The vertical axes show the
  relative differences $|S^{bc}_{\psi^z_1} - S^{bc}_{\psi_0}|/|S^{bc}_{\psi_0}|$
  with $b=c=z$ (left panels) and $b=c=x$ (right panels).
  Along the horizontal axis, the number of spins in the open
  direction ($N_x$) is varied. The different curves correspond to configurations
  with different $N_y$. The insets show for which sites the correlation
  functions were computed. The sites in the central column of the insets
  correspond to the middle of the cylinder in the $x$-direction.
  The relative difference decreases exponentially in $N_x$.}
\end{figure}

We now discuss the question if the states $\psi^a_1$ can be
considered as edge states. If so, then the local properties of
$\psi^a_1$ and $\psi_0$ in the bulk should be the same.  Since these
are encoded in the spin correlation functions, we compared the
nearest-neighbor two-point correlators in the bulk for different
system sizes.  The relative differences
\begin{align}
  \label{eq:relative-differences-uz-psi0-vs-psi0}
  \left|\frac{S^{bc}_{\psi^a_1}(i_x, j_x, \Delta_y) - S^{bc}_{\psi_0}(i_x, j_x, \Delta_y)}{S^{bc}_{\psi_0}(i_x, j_x, \Delta_y)} \right|
\end{align}
are shown in Fig.~\ref{fig:psi0-vs-uz-psi0} for $a=b=c=z$ (left
panels) and $a=z, b=c=x$ (right panels).  Correlation functions for
other choices of $a, b$, and $c$ either vanish or can be reduced to
these due to the SU(2) invariance of $\psi_0$.  In the upper panels,
the correlations along the $y$-direction are shown and in the
panels of the lower two rows along the $x$-direction.  We find that
the relative differences approach zero exponentially as a function of
$N_x$. Even though the differences tend to be larger for smaller
$N_y$, they are still exponentially suppressed as $N_x$ is
increased. This is an indication that the wave functions $\psi^a_1$
indeed describe edge states compared to $\psi_0$ as the thermodynamic
limit in the open direction is taken.

\begin{figure}[htb]
\centering
\includegraphics[width=1.0\linewidth]{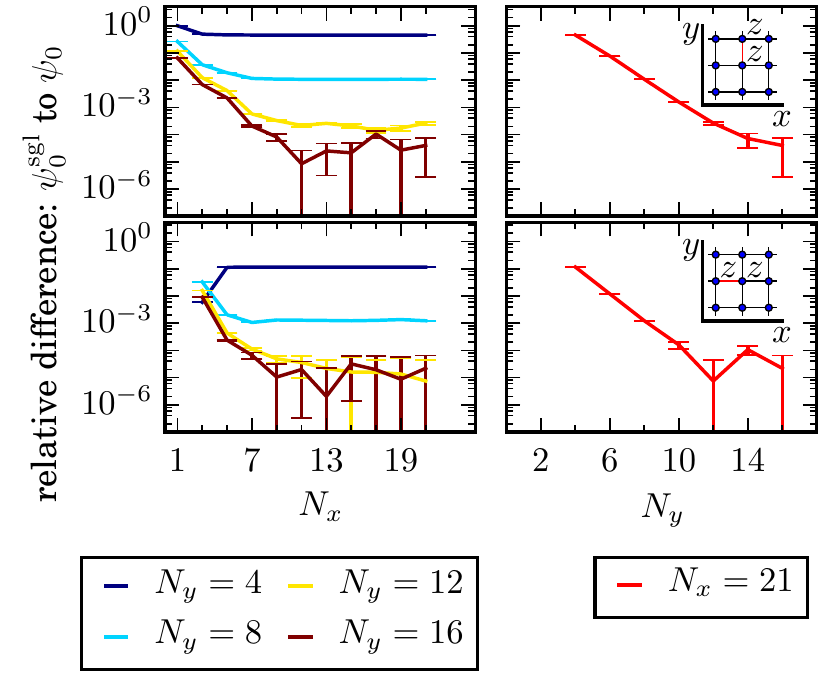}
\caption{\label{fig:psi0-vs-psi0sgl} (Color online)
  Comparison of nearest-neighbor bulk correlations in
  $\psi^{\text{sgl}}_0$ and $\psi_0$. The vertical axes show the relative
  differences $|S^{zz}_{\psi^{\text{sgl}}_0} - S^{zz}_{\psi_0}|/|S^{zz}_{\psi_0}|$.
  In the left panels, $N_x$ is varied along the horizontal axes
  and the different curves correspond to different choices of $N_y$.
  On the right panels, $N_y$ varies along the horizontal axes and $N_x$
  is fixed. For $N_x$ large enough, the differences tend
  to zero exponentially as a function of $N_y$.}
\end{figure}

Our results for the comparison between $\psi^{\text{sgl}}_0$ and $\psi_0$
are shown in Fig.~\ref{fig:psi0-vs-psi0sgl}. Since both $\psi_0$
and $\psi^{\text{sgl}}_0$ are singlets, it is enough to compare the $zz$-correlations.
Furthermore, the correlations in the positive and the negative $x$-direction
are the same in the middle of the cylinder since $\psi_0$ and $\psi^{\text{sgl}}_0$ are symmetric
under the inversion, cf. Sec.~\ref{sec:global-properties}.
In contrast to $\psi^a_1$, we find that the thermodynamic limit in the
$x$-direction is not enough for the differences to vanish. Rather, we
observe that the differences become stationary if $N_y$ is held fixed and
$N_x$ increased. As shown in the right panels of Fig.~\ref{fig:psi0-vs-psi0sgl},
the differences do, however, tend to zero
exponentially as a function of $N_y$ if $N_x$ is chosen large enough.

\subsection{States at a higher level}
In the previous subsection, the states at level one in
current operators were considered. We also compared spin
correlations in $\psi^a_n$ to those in $\psi_0$ for higher values
of $n$.  For very large mode numbers $n$, only the terms at the edge
contribute to the sum in $u^a_{-n}$. To see this, let us consider
\begin{align}
  \label{eq:limit-of-u-for-high-modes-number}
  u^a_{-n - m N_y} &= \sum_{j=1}^N \frac{1}{z_j^{n + m N_y}} t^a_j\\
  &\propto \sum_{j_x=1}^{N_x} e^{-\frac{2 \pi}{N_y} (n + m N_y) j_x} \sum_{j_y=1}^{N_y} e^{-\frac{2 \pi i}{N_y} n j_y} t^a_{j_x, j_y}.
\end{align}
For large values of $m$, the terms with $j_x > 1$ are
exponentially suppressed with respect to those that have $j_x = 1$.
We denote the corresponding states with
one current operator by $\chi^a_n$:
\begin{align}
  \label{eq:limit-only-edge}
  | \chi^a_n \rangle
  = \lim_{m \to \infty} | \psi^a_{n + m N_y} \rangle
  \propto \sum_{j_y=1}^{N_y} e^{-2 \pi i \frac{n j_y}{N_y}} t^a_{1, j_y} | \psi_0 \rangle.
\end{align}

\begin{figure}[htb]
\centering
\includegraphics[width=1.0\linewidth]{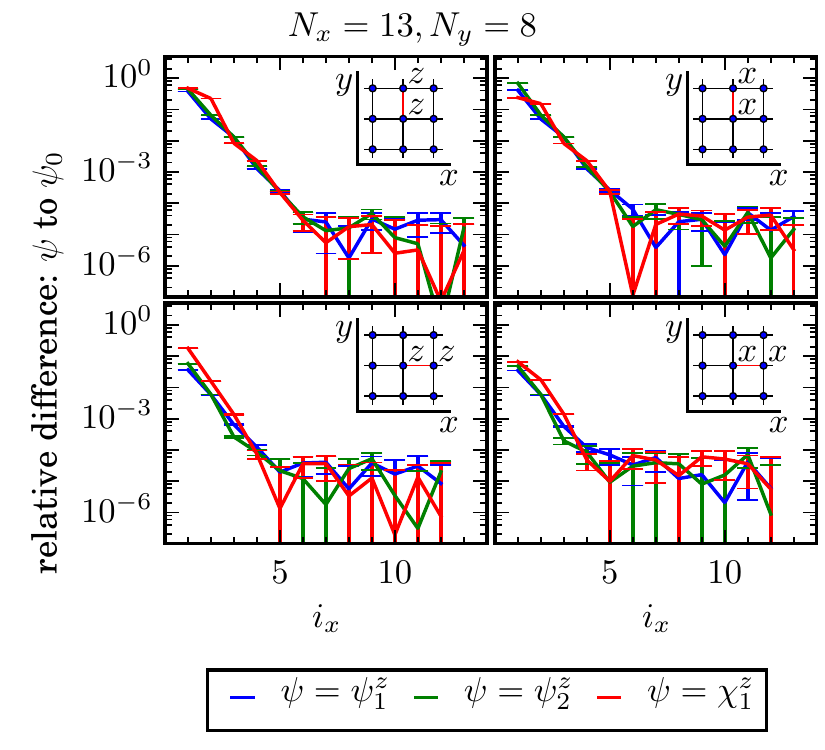}
\caption{\label{fig:comparison-to-psi0-for-higher-levels}
  (Color online) Relative difference $|S^{ab}_{\psi} -
S^{ab}_{\psi_0}|/|S^{ab}_{\psi_0}|$ in nearest-neighbor correlators
for $\psi \in \{\psi^z_1, \psi^z_2, \chi^z_1\}$
[cf. Eq.~\eqref{eq:limit-only-edge} for the definition of $\chi^a_1$].
The position $i_x$ in the $x$-direction is varied along the horizontal
axis.  The plots in the left panels have $a=b=z$ and those in the
right panels $a=b=x$.  In the upper panels, the correlations along the
$y$-direction are shown [$j_x = i_x, \Delta_y=1$] and the lower panels
correspond to correlations along the $x$-direction [$j_x = i_x + 1$,
$\Delta_y = 0$].
}
\end{figure}

Fig.~\ref{fig:comparison-to-psi0-for-higher-levels} shows the
difference in nearest-neighbor correlations along the
$y$-direction relative to $\psi_0$ for $N_x=13$ and $N_y=8$.
The three curves correspond to the
states $\psi^z_1, \psi^z_2$ and $\chi^z_1$. As the position in the
open direction is increased, the differences vanish exponentially for
all three states. We note that the differences are large at the left edge ($i_x = 1$)
and small at the right edge ($i_x = 13$).
This agrees with the expectation that the operators
$u^a_{-n}$ are localized at the left edge. In contrast to the state
$\psi^{\mathrm{sgl}}_0$, the states $\psi^a_n$ perturb $\psi_0$ only
at one edge and their behavior is therefore expected to approach
that of $\psi_0$ at the other edge. The results of
Fig.~\ref{fig:comparison-to-psi0-for-higher-levels} provide an
indication that $\psi^a_n$ describe edge states also for $n > 1$.

We note, however, that the linear span of $\psi^a_n$
for $n \in \{1, \dots, N-1\}$ contains not only edge states. The states $\psi^a_n$ can
even be linearly combined so that $\psi_0$ is perturbed at an
arbitrary position $j$:
\begin{align}
  \label{eq:perturbartion-at-arbitrary-position}
  t^a_j | \psi_0 \rangle = \sum_{n=2}^N \left(\mathcal{Z}^{-1}\right)_{j n} | \psi^a_{n-1} \rangle,
\end{align}
where $\mathcal{Z}$ is the matrix defined in
Eq.~\eqref{eq:mat-spin-operator-basis-transformation}.
This observation can be understood from the fact that two
states $\psi^a_m$ and $\psi^a_n$ are not necessarily orthogonal
if $m - n = 0$ modulo $N_y$. In this case, $\psi^a_m$
and $\psi^a_n$ have the same momentum in the $y$-direction,
as discussed in Sec.~\ref{sec:global-properties}. The linear combination
\begin{align}
  \label{eq:linear-combination-without-edge-contribution}
  &| \psi^a_n \rangle - e^{-2  \pi \frac{N_x - 1}{2}} | \psi^a_{n + N_y} \rangle \notag \\
  &= \sum_{j_x=1}^{N_x} (1 - e^{- 2 \pi (j_x - 1)}) \sum_{j_y=1}^{N_y} \frac{1}{z_{j_x, j_y}^n} t^a_{j_x, j_y} | \psi_0 \rangle,
\end{align}
for example, receives no contribution from spin operators at the left
edge ($j_x = 1$).  Even though both $\psi^a_n$ and $\psi^a_{n + N_y}$
are perturbed from $\psi_0$ mostly at $j_x = 1$, this is not the case
for the difference of
Eq.~\eqref{eq:linear-combination-without-edge-contribution}.

\subsection{Inner products of states from current operators}
In this subsection, we discuss the relation of inner products between
the states $\Jpsi$ on the level of the spin system and CFT inner
products between states $\JCFT$.  For edge states in the continuum
that are constructed from descendant states of a CFT,
the authors of Ref.~\onlinecite{Dubail2012a} come to the
remarkable conclusion that, in the thermodynamic limit and under the
assumption of exponentially decaying correlations in the bulk, the inner products
between edge states are the same as the inner products between CFT
states. We now consider inner products between states constructed
from current operators to test if a similar correspondence
holds for the lattice states $\Jpsi$ and the CFT states
they are  constructed from.
The spin system inner products that we consider are given by
\begin{align}
\label{eq:Jpsi-inner-products}
  &R^{k + k'} \frac{\langle \Jpsi | \Jpsib \rangle}{\langle \psi_0 | \psi_0 \rangle} \notag \\
  &\equiv R^{k + k'}
    \frac{\langle \psi_0 |
    \left(u^{a_1}_{-n_1}\right)^\dagger \dots \left(u^{a_l}_{-n_l}\right)^\dagger
    u^{b_{l'}}_{-m_l'} \dots u^{b_1}_{-m_1}
    | \psi_0 \rangle}
    {\langle \psi_0 | \psi_0 \rangle},
\end{align}
where $N$ is the number of spins,
\begin{align}
  \label{eq:definition-of-R}
R = \min_{j\in\{1,\dots,N\}} |z_j| = e^{-\frac{\pi}{N_y} (N_x -1)}
\end{align}
is the minimal absolute value of the positions, $k =
\sum_{j=1}^l n_j$, and $k' = \sum_{j=1}^{l'} m_j$.
The factor $R^{k + k'}$ accounts for the scaling of the
operators $u^{a}_{-n}$ with respect to a rescaling of the
positions. The minimal value is chosen because the
operators
\begin{align}
  \label{eq:constribution-from-edge-in-u}
  u^a_{-n} = \sum_{j=1}^N \frac{t^a_j}{(z_j)^n}
\end{align}
have the highest contribution at the edge with $|z_j| = R$.

We compare the inner products of the lattice system~\eqref{eq:Jpsi-inner-products}
to the CFT inner products
\begin{align}
  \label{eq:CFT-inner-products}
  \langle 0 | J^{a_1}_{n_1} \dots J^{a_l}_{n_l} J^{b_{l'}}_{-m_{l'}} \dots J^{b_1}_{-m_1} | 0 \rangle.
\end{align}
If a correspondence similar to that of Ref.~\onlinecite{Dubail2012a}
also holds for lattice states, then the expressions of
Eq.~\eqref{eq:Jpsi-inner-products} should approach those of
Eq.~\eqref{eq:CFT-inner-products} in the thermodynamic limit.

Note that the inner products of the spin system are hard to evaluate for large
system sizes, whereas the CFT inner product can be easily computed
using the Kac-Moody algebra \eqref{eq:Kac-Moody-current-algebra}.  On
the level of the spin system, the insertion of current operators
corresponds to an application of spin operators to $\psi_0$
[cf. Eq.~\eqref{eq:jpsi-from-psi0}]. Therefore, the inner products can
be determined numerically using a Monte Carlo method if the number of
current operators is small.

We calculated the inner products for the states $\psi^a_1$,
$\psi^a_2$ and $\psi^{b\phantom{,}b}_{1,1}$, which are all nonzero states at levels one and two.
For these states, inner products
between different states vanish because they have either a different
spin or a different momentum. It is thus sufficient to compare the
norm squared of a state to the norm squared of the corresponding
CFT state, as summarized in the following table:
\begin{center}

\begin{tabular}{lll}
                   Spin state &                           CFT state &                                           Norm squared of CFT state \\
\hline
                   $\psi^a_1$ &          $J^a_{-1} | 0 \rangle $ & $\langle J^a_{1} J^a_{-1} \rangle  = \frac{1}{2}$ (no sum over $a$) \\
                   $\psi^a_2$ &          $J^a_{-2} | 0 \rangle $ &           $\langle J^a_{2} J^a_{-2} \rangle  = 1$ (no sum over $a$) \\
 $\psi^{b\phantom{,}b}_{1,1}$ & $J^b_{-1} J^b_{-1} | 0 \rangle $ &   $\langle J^c_{1} J^c_{1} J^b_{-1} J^b_{-1} \rangle = \frac{9}{2}$ 
\end{tabular}
\end{center}

\begin{figure}[htb]
\centering
\includegraphics[width=1.0\linewidth]{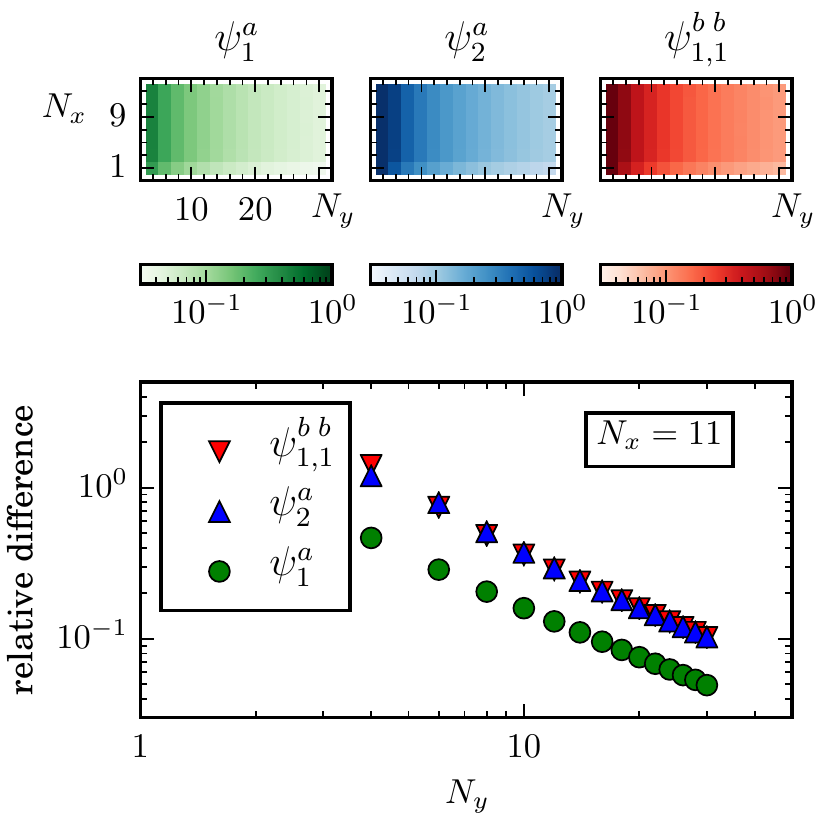}
\caption{\label{fig:inner-products} (Color online) Inner product of
  spin system states: Relative difference to the CFT expectation
  [cf. Eq.~\eqref{eq:relative-difference-to-inner-products}].
  The colors
  correspond to the states $\psi^{b\protect\phantom{,}b}_{1,1}$ (red),
  $\psi^{a}_2$ (blue) and $\psi^{a}_1$ (green). The upper panels show the
  relative difference in a color plot as a function of $N_x$ and $N_y$.
  We observe a very weak dependence on $N_x$ for $N_x \ge 3$. The lower panel
  shows the dependence on $N_y$ for $N_x = 11$. For large enough $N_y$,
  the data are consistent with a power-law behavior with an exponent
  of approximately $-1.1$. Monte Carlo error bars are not plotted because
  they are barely visible on the chosen scale. The maximal relative error
  of the shown data points is 0.31~\%.
}
\end{figure}

In Fig.~\ref{fig:inner-products}, our numerical results are shown for
the relative difference
\begin{align}
  \label{eq:relative-difference-to-inner-products}
  \left|\frac{R^{2k} \dfrac{\langle \psi | \psi \rangle}{\langle \psi_0 | \psi_0 \rangle}
  - \langle \psi_{\text{CFT}} | \psi_{\text{CFT}} \rangle}
  {\langle \psi_{\text{CFT}} | \psi_{\text{CFT}} \rangle}\right|
\end{align}
as a function of the system size.  Here, $\psi \in \{\psi^a_1,
\psi^a_2, \psi^{b\phantom{,}b}_{1,1}\}$ is one of the spin states,
$k=1$ for $\psi = \psi^a_1$, $k=2$ for $\psi \in \{\psi^a_2,
\psi^{b\phantom{,}b}_{1,1}\}$, and $\psi_{\mathrm{CFT}}$ is the CFT
state corresponding to $\psi$.

For a given system size, we observe a
smaller difference for the states at level $k=1$ than for those at level
$k=2$.  The computed inner products approach the CFT expectation if $N_y$ is
increased.  The dependence on the number of spins in the $x$-direction
is, however, very weak for $N_x \ge 3$. In particular, the CFT result
is not approached if $N_x$ is increased and $N_y$ kept fixed. For
large enough $N_y$, our data suggest that the spin system inner
products approach the CFT result with a power law in $N_y$.

\section{Local model Hamiltonian}
\label{sec:local-model-hamiltonians}
In the previous section, we provided numerical evidence that the
states with one current operator insertion represent edge states with
respect to $\psi_0$.  In this section, we study a set of local
Hamiltonians on the cylinder. For a suitable choice of parameters, the
ground state of the corresponding Hamiltonian has a good overlap with
$\psi_0$ and some of its low-energy excited states are well approximated
by $\psi^a_1$, the states with one current operator of
order one.

We study the local Hamiltonians\cite{Nielsen2013a}
\begin{align}
  \label{eq:definition-of-local-Hamiltonians}
  H &= J_2 \sum_{\langle i, j\rangle} t^a_i t^a_j + J'_2 \sum_{\llangle i, j \rrangle} t^a_i t^a_j
      + J_3 \sum_{\langle i, j, k\rangle_\circlearrowleft} \vep_{abc} t^a_i t^b_j t^c_k.
\end{align}
In these sums, the sites lie on a square lattice, $\langle i, j
\rangle$ denotes all nearest neighbors, $\llangle i, j\rrangle$ all
next-to-nearest neighbors, and $\langle i, j,
k\rangle_\circlearrowleft$ all triangles of nearest neighbors for
which $i, j$, and $k$ are oriented counter-clockwise.  It was shown in
a previous study\cite{Nielsen2013a} that the ground state of $H$ on
the plane (open boundary conditions in both directions) and on the
torus has a good overlap with the KL state for a range of parameters
$J_2, J'_2$, and $J_3$. Here, we study $H$ on a cylinder of size $N_x
\times N_y$, where $N_x$ denotes the number of sites in the open
direction and $N_y$ the number of sites in the periodical direction.
In the following, we parametrize $H$ in terms of two angles $\theta_1$ and $\theta_2$:
\begin{align}
  \label{eq:parametrization-of-local-Hamiltonians}
  J_2 &= \cos\left(\theta_1 \right) \cos\left(\theta_2\right), \notag \\
  J'_2 &= \sin\left(\theta_1\right) \cos\left(\theta_2\right),\notag \\
  J_3 &= \sin\left(\theta_2\right).
\end{align}

For $N_x = 5$ and $N_y = 4$, we studied the overlap between
$\psi_0$ and the ground state $\psi_G$ of $H$ as a function of
$\theta_1$ and $\theta_2$ using an exact numerical diagonalization
method. We also computed the overlap of the states with one
current operator insertion at level one $\psi^a_1$ and the first
excited states $\psi^m_E$ of $H$ that have spin one and the same
momentum in the $y$-direction as $\psi^a_1$. Here, $m \in \{-1, 0, 1\}$ denotes
the $T^z$ eigenvalue of $\psi^m_E$.

\begin{figure}[htb]
\centering
\includegraphics{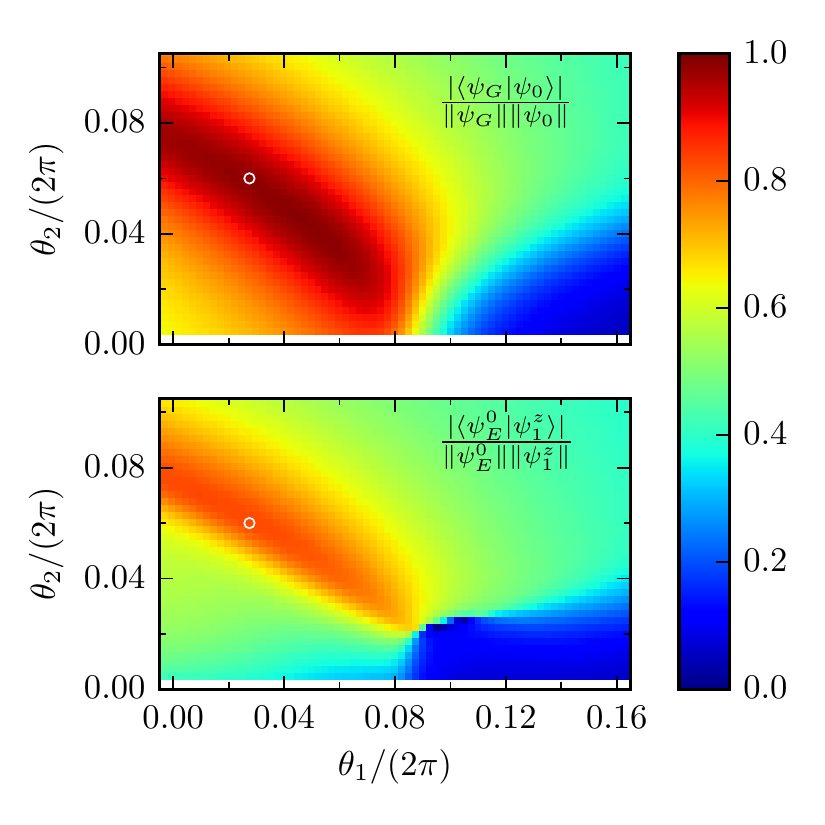}
\caption{\label{fig:overlaps-as-function-of-angles}
  (Color online)
  Overlaps of states constructed from CFT and eigenstates of the local Hamiltonian $H$
of Eq.~\eqref{eq:definition-of-local-Hamiltonians} for $N_x = 5$ and
$N_y = 4$. The angles $\theta_1$ and $\theta_2$ parametrize the
coupling constants of $H$ according to
Eq~\eqref{eq:parametrization-of-local-Hamiltonians}.  In the upper
panel, the overlap $\olap{\psi_G}{\psi_0} \equiv |\langle \psi_G| \psi_0\rangle|/(\|\psi_G\| \|\psi_0\|)$ between
$\psi_0$ and the ground state $\psi_G$ of $H$ is plotted. The lower
panel shows the overlap between $\psi^z_1$ and the first excited state
$\psi^0_E$ of $H$ with the same spin and $y$-momentum as $\psi^z_1$ [spin
one, $T^z = 0$, momentum $3 / (8 \pi)$]. The point marked with an
open circle has $\theta_1 = \thetaoneopt$ and $\theta_2 =
\thetatwoopt$ and the highest combined overlap of $\compol \approx 1.2858$.}
\end{figure}

We denote the overlap between two states $\phi_1$ and $\phi_2$
as
\begin{align}
  \label{eq:definition-of-overlap}
  \olap{\phi_1}{\phi_2} = \frac{|\langle \phi_1 | \phi_2 \rangle|}{\|\phi_1\| \|\phi_2\|},
\end{align}
where $\|\phi\| = \sqrt{\langle \phi | \phi \rangle}$ is the norm of a state.
In Fig.~\ref{fig:overlaps-as-function-of-angles}, the overlaps
$\olap{\psi_G}{\psi_0}$ and $\olap{\psi^0_G}{\psi^z_1}$ are shown as a function of the parameters of the
Hamiltonian.  Due to SU(2) invariance, it is sufficient to consider
the overlap between the states $\psi^0_E$ and $\psi^z_1$, which both
have $T^z=0$:
\begin{align}
  \label{eq:su2-invariance-for-overlaps}
  |\langle \psi^{1}_E | \psi^{+}_1 \rangle|
  = |\langle \psi^{-1}_E | \psi^{-}_1 \rangle|
  = |\langle \psi^{0}_{E} | \psi^{z}_1 \rangle|,
\end{align}
where $\psi^{\pm}_1 \equiv \psi^{x}_1 \pm i \psi^{y}_1$. The best value for the combined overlap $\compol$
was obtained for the angles $\theta_1 = \thetaoneopt$ and $\theta_2 = \thetatwoopt$:
\begin{center}

\begin{tabular}{llll}
 $\olap{\psi_G}{\psi_0}$ & $\olap{\psi^0_E}{\psi^z_1}$ & $\olap{\psi_G}{\psi_0}^{\frac{1}{N}}$ & $\olap{\psi^0_E}{\psi^z_1}^{\frac{1}{N}}$ \\
\hline
                $0.9829$ &                    $0.8289$ &                              $0.9991$ &                                  $0.9907$ 
\end{tabular}
\end{center}
Since the size of the Hilbert space grows exponentially with the system size $N$, the
overlaps are expected to scale exponentially in $N$. By taking the $N$th root,
one obtains a measure for the overlap per site, which takes into
account this exponential scaling. Notice that the overlap per site
is higher than 99 \% for both the ground and the excited state.

\begin{figure}[htb]
\centering
\includegraphics{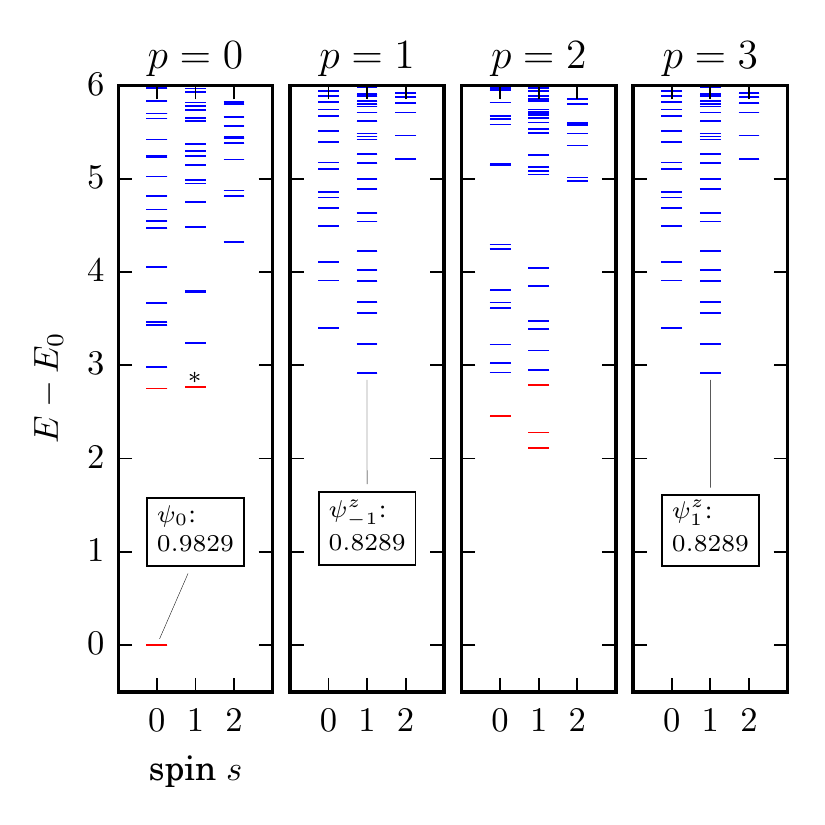}
\caption{\label{fig:spectrum}
  (Color online) Low energy spectrum of the Hamiltonian $H$ of
Eq.~\eqref{eq:definition-of-local-Hamiltonians} for $N_x = 5, N_y = 4,
\theta_1 = \thetaoneopt$, and $\theta_2 = \thetatwoopt$. The ground state energy of $H$
is $E_0 \approx -24.11$. The four
panels correspond to the four sectors of $y$-momentum $p/(2 \pi N_y)$
with $p \in \{0, 1, 2, 3\}$. Each shown level has a degeneracy of $2 s
+ 1$ corresponding to the values of $T^z$. The labels show the ansatz state
constructed from CFT and the value for its overlap with the
corresponding eigenstate of $H$. The 8 energies shown in red are those that
are smaller than the lowest energy in the $p=1$ and $p=3$
sectors. [At the level marked with an asterisk ($\ast$), there are actually
two energies with a splitting of approximately $1.211 \times 10^{-3}$. This is not
visible on the scale of the plot.]}
\end{figure}

The low-energy spectrum of $H$ for the parameters
with the best overlaps is plotted in Fig.~\ref{fig:spectrum}. We find 8 energies
below the energy of $\psi^m_E$.
The spectra plotted in Fig.~\ref{fig:spectrum} are separated into
sectors of different $y$-momentum $p / (2 \pi N_y) = p / (8 \pi)$
with $p \in \{0, 1, 2, 3\}$.
Note that the states $\psi^m_E$ are the first excited states with $p=3$.
The spectra for the momenta for $p=1$ and $p=3$ are the same
because $H$ is invariant under the inversion
operator $\mathcal{I}$ introduced in Sec.~\ref{sec:global-properties}:
\begin{align}
  \label{eq:invariance-of-H-under-I}
  \mathcal{I}^{-1} H \mathcal{I} = H.
\end{align}
The relation
\begin{align}
  \label{eq:I-transformation-of-Ty}
  \mathcal{I}^{-1} \mathcal{T}_y \mathcal{I} = \mathcal{T}_y^{-1}
\end{align}
between $\mathcal{I}$ and translation operator in the $y$-direction $\mathcal{T}_y$
follows directly from their definition
(cf. Appendix~\ref{sec:appendix-translation-operator} and~\ref{sec:appendix-inversion-operator}).
Therefore, if $| \psi \rangle$ is an eigenstate of $H$ with momentum $p / (2 \pi N_y)$,
then $\mathcal{I} | \psi \rangle$ is also an eigenstate with momentum $(N_y - p)/(2 \pi N_y)$.
This means that for $| \psi^m_E \rangle$ with $p=3$, there is a corresponding eigenstate
$\mathcal{I} | \psi^m_E \rangle$ with $p=1$, which satisfies
\begin{align}
  \label{eq:overlaps-in-corresponding-sectors-of-momentum}
  |\langle \psi^0_E | \psi^z_1 \rangle | &= |\langle \mathcal{I} \psi^0_E | \psi^z_{-1} \rangle|.
\end{align}
Here, $| \psi^z_{-1} \rangle = \mathcal{I} | \psi^z_1 \rangle$ is the state obtained by inserting the
current operator at $z_\infty = \infty$ instead of $z_0 = 0$
[cf. Eq.~\eqref{eq:definition-of-Jpsi-inf}].

Our results show that the state $\psi_0$ and the states
with one current operator of order one are good approximations
of low-energy eigenstates of $H$ for $N_x = 5$, $N_y = 4$, 
$\theta_1 = \thetaoneopt$, and $\theta_2 = \thetatwoopt$.
This raises the question if further eigenstates of $H$ are
effectively described by states constructed as CFT correlators.
We also computed the overlaps of eigenstates of $H$ with some
additional states
constructed from current operators for higher orders in current
operators. At level two in current operators, the overlaps
with the first excited states that have the same spin and momentum
as our ansatz states are given by $0.5486$ for $\psi^a_2$ ($0.9704$ per site)
and $0.3301$ for $\psi^{b\phantom{,}b}_{1, 1}$ ($0.9461$ per site).

In this Section, we considered a local model and computed overlaps
between ansatz states constructed from CFT and eigenstates of the
model Hamiltonian. In Appendix~\ref{sec:exact-parent-hamiltonians}, we
also derive exact, SU(2) invariant parent Hamiltonians for some of the
states constructed from current operators. These Hamiltonians are
nonlocal with up to four-body interactions.

\section{Conclusion}
\label{sec:conclusion}
This work studies trial wave functions for lattice FQH states
constructed as chiral correlators of the SU(2)$_1$ WZW CFT.
To each CFT state, characterized by a sequence of current
operators, we associated a corresponding state $\Jpsi$ of the lattice
system.  For continuous systems, analogous states constructed from CFT
were proposed as FQH edge states
previously\cite{Wen1994, Dubail2012a}. The fact that we work on the
lattice allowed us to apply Monte Carlo techniques to test a central
expectation for edge states: That the local, bulk properties of
different edge states should be the same.

For a system on the cylinder, we compared spin correlation
functions in the states with one current operator ($\psi^a_1$) to the
state with no current operators ($\psi_0$).  Our numerical results show
that the nearest-neighbor bulk correlations approach each
other exponentially as the number of spins in the open direction ($N_x$)
is increased.  On the other hand, the states $\psi^a_1$ and
$\psi_0$ are different globally since their spin and momentum are
different. This supports the assumption that they describe edge states.

We compared inner products of lattice states at levels one
and two in current operators to CFT inner products of
the corresponding descendant states.
For large enough $N_y$ (periodical direction), the computed inner
products approach the CFT expectation with a power law in $N_y$. This
suggests that there is a correspondence between inner products of
states $\Jpsi$ and CFT inner products in the thermodynamic limit. Such
a correspondence was found for continuous wave functions in
Ref.~\onlinecite{Dubail2012a}.

Furthermore, we compared nearest-neighbor bulk correlations in
$\psi^{\mathrm{sgl}}_0$ to those in $\psi_0$, where $\psi^{\mathrm{sgl}}_0$ is
the singlet component of the state obtained by insertion of two extra
primary fields. In contrast to $\psi^a_1$, we find that the
correlations do not approach each other if the thermodynamic limit is
taken only in the open direction. However, if $N_x$ is chosen large
enough, the difference in correlation functions vanishes exponentially
as a function of $N_y$.

We showed by an exact diagonalization that $\psi_0$ has a good overlap
with the ground state of a local Hamiltonian and $\psi^a_1$ with the
first excited states that have the same spin and momentum as
$\psi^a_1$. This could be an indication that further
low-energy excitations of
that local Hamiltonian are edge states described by the SU(2)$_1$ WZW CFT.
It would be interesting to investigate this relation in more detail
for larger system sizes and different topologies.

We showed that the complete Hilbert space is covered by the linear
span of the states $\Jpsi$ and, therefore, only a subset of these
states are edge states. For the states with one current operator, we argued
that not all linear combinations of states $\psi^a_m$ describe edge modes
because states with the same $y$-momentum can be non-orthogonal. It is
possible to restrict the space of states to an orthogonal subset
given by $\psi^a_m$ with $m \in \{1, \dots, N_y\}$.

Taking the limit of large mode numbers could be another
possibility of removing bulk states for the linear span
of $\Jpsi$. More precisely,
one can replace $n_i$ by $n_i + m_i N_y$ and then take
$m_i \to \infty$. In this limit, the sum in the operators $u^a_{-n_i -
  m_i N_y}$ only extends over the edge sites because all other positions
are exponentially suppressed.
The fact that this class of states (and also linear combinations of
such states) is obtained from $\psi_0$ by application of edge spin
operators only, suggests that their complete span represents edge
states.  For one of these states, $\chi^a_1$, we did numerical tests
that indeed indicated that $\chi^a_1$ is an edge state.

\begin{acknowledgments}
  We acknowledge funding from the EU Integrated Project SIQS, FIS2012-33642,
  the Comunidad de Madrid grant QUITEMAD+ S2013/ICE-2801 (CAM),
  the Severo Ochoa Program, and
  the Villum Foundation.
\end{acknowledgments}

\appendix
\section{Translation and inversion of states on the cylinder}
\label{sec:appendix-derivation-of-global-properties}
\subsection{Transformation under a permutation of the spins}
Both the translation operator $\mathcal{T}_y$ and the inversion operator $\mathcal{I}$
act on a product state as a permutation of the spins. Such a permutation
operator $\mathcal{O}_\tau$ is defined for the permutation $\tau$ of
$N$ elements as
\begin{align}
  \label{eq:permutation-tau}
  \mathcal{O}_\tau |s_1, \dots, s_N \rangle &= |s_{\tau(1)}, \dots, s_{\tau(N)}\rangle.
\end{align}

The action of $\mathcal{O}_\tau$ on $\psi_0$ and $\psi^{s_0, s_\infty}_0$ can be
rewritten in terms of a permutation of the positions $z_i$, which will
facilitate our calculations for $\mathcal{T}_y$ and $\mathcal{I}$.
Our derivation of this transformation rule follows Ref.~\onlinecite{Nielsen2013d}.

We consider the wave function
\begin{align}
\label{eq:psi-0-tilde}
\tilde{\psi}^{z_1, \dots, z_N}_0(s_1, \dots, s_N) &= \delta_{\mathbf{s}} \chi_{\mathbf{s}} \prod_{i < j}^{N} (z_i - z_j)^{(s_i s_j+1) / 2},
\end{align}
which is equivalent to $\psi_0$ because it only differs by a spin-independent constant
[cf. Eq.~\eqref{eq:vertex-op-correlator}].
We have also explicitly written out the parametric dependence on the
positions $z_i$. Similarly, the wave function
\begin{align}
\label{eq:psi-0inf-equivalent}
  &\tilde{\psi}^{s_0, s_\infty, z_1, \dots, z_N}_0(s_1, \dots, s_N) \notag\\
  &=\delta_{\bar{\mathbf{s}}} (-1)^{(1-s_\infty)/2} \chi_{\mathbf{s}} \prod_{n=1}^{N} z_n^{(s_0 s_n + 1) / 2}\notag \\
  &\quad  \times   \prod_{n < m}^N (z_n - z_m)^{(s_n s_m + 1) / 2}
\end{align}
is equivalent to $\psi^{s_0, s_\infty}_0$.
Let us first calculate the transformation of $\tilde{\psi}^{z_1,
\dots, z_N}_0$ under a simultaneous permutation of both the spins and
the coordinates. Since every permutation can be decomposed into a
series of transpositions, we consider the case that $\tau$ is a
transposition:
\begin{align}
  \label{eq:definition-of-transposition}
  \tau(i) &=
            \begin{cases}
              i, &\text{if } i \notin \{m, n\},\\
              n, &\text{if } i = m,\\
              m, &\text{if } i = n,
            \end{cases}
\end{align}
for $m, n \in \{1, \dots, N\}$ and $m < n$.
It follows that
\begin{align}
  \label{eq:derivation-of-combined-transposition}
  &\frac{\tilde{\psi}^{z_{\tau(1)}, \dots, z_{\tau(N)}}_0(s_{\tau(1)}, \dots, s_{\tau(N)})}{\tilde{\psi}^{z_1, \dots, z_N}_0(s_1, \dots, s_N)} \notag \\
  &= \underbrace{(-1)^{(n - m) (s_m - s_n)/2}}_{\text{transformation of } \chi_{\mathbf{s}}} \prod_{\substack{i < j,\\ \tau(i) > \tau(j)}}^{N} (-1)^{(s_{\tau(i)} s_{\tau(j)} + 1)/2} \notag \\
  &= (-1)^{(n - m) (s_m - s_n)/2} (-1)^{(s_m  s_n + 1)/2} \notag \\
  &\quad \times \prod_{j = m+1}^{n-1} (-1)^{(s_j (s_m + s_n) + 2)/2} \notag \\
  &= -1.
\end{align}
Therefore, if $\tau$ is a general permutation corresponding to $\mathcal {N}_\tau$ subsequent transpositions,
\begin{align}
  \label{eq:transformation-of-positions-and-spins}
  &\tilde{\psi}^{z_{\tau(1)}, \dots, z_{\tau(N)}}_0(s_{\tau(1)}, \dots, s_{\tau(N)}) \notag \\
  &= \mathrm{sign}(\tau) \tilde{\psi}^{z_1, \dots, z_N}_0(s_1, \dots, s_N),
\end{align}
where $\mathrm{sign}(\tau) = (-1)^{\mathcal{N}_\tau}$ is the signature of the permutation.
Substituting $s_i$ by $s_{\tau^{-1}(i)}$ in Eq.~\eqref{eq:transformation-of-positions-and-spins}, we arrive at the final
transformation rule:
\begin{align}
  \label{eq:final-transformation-psi0-under-permutation}
  &\tilde{\psi}^{z_1, \dots, z_N}_0(s_{\tau^{-1}(1)}, \dots, s_{\tau^{-1}(N)})\notag \\
  &= \mathrm{sign}(\tau) \tilde{\psi}^{z_{\tau(1)}, \dots, z_{\tau(N)}}_0(s_1, \dots, s_N).
\end{align}
The transformation under a permutation of the spins can therefore be calculated
by considering the corresponding transformation of the coordinates and taking into account
the signature of the permutation.

We note that Eq.~\eqref{eq:final-transformation-psi0-under-permutation} is not valid
for the original wave function $\psi_0$ but only for $\tilde{\psi}^{z_1, \dots, z_N}_0$,
which differs from $\psi_0$ by a factor depending on $z_i$. However, this factor
does not depend on the spins. Therefore, if
$\tilde{\psi}^{z_1, \dots, z_N}_0$ is an eigenstate of $\mathcal{O}_\tau$, then this
is also the case for $\psi_0$.

Compared to $\tilde{\psi}_0$, there are some an additional factors present
the wave function $\tilde{\psi}^{s_0, s_\infty, z_1, \dots, z_N}_0$. Since these
are invariant under a permutation of both the spins and the coordinates,
a formula analogous to Eq.~\eqref{eq:final-transformation-psi0-under-permutation} holds for
$\tilde{\psi}^{s_0, s_\infty, z_1, \dots, z_N}_0$.

\subsection{Translation in the periodical direction}
\label{sec:appendix-translation-operator}
The translation operator $\mathcal{T}_y$ is defined through
the permutation $\tilde{\mathcal{T}}_y$:
\begin{align}
\label{eq:definition-translation-as-permutation}
\tilde{\mathcal{T}}_y\left(i_x, i_y\right) &=
\begin{cases} (i_x, i_y + 1), &\text{if } i_y \neq N_y ,\\
(i_x, 1), &\text{if } i_y = N_y,
   \end{cases}
\end{align}
where $i_x$ is the $x$-component and $i_y$ the $y$-component
of an index $i$.

The signature of this permutation is given by
\begin{align}
  \label{eq:signature-of-Ty}
\mathrm{sign}(\tilde{\mathcal{T}}_y) = (-1)^{N_x (N_y - 1)} = (-1)^{N_x},
\end{align}
where we used that $N=N_y N_x$ is even.
 In terms of the positions, the transformation
corresponds to a multiplication by a phase,
$z_{\tilde{\mathcal{T}}_y(j)} = e^{2 \pi i / N_y} z_j$. Therefore,
\begin{align}
\label{eq:transformation-unter-translation}
 &\tilde{\psi}^{z_1, \dots, z_N}_0(s_{\tilde{\mathcal{T}}_y^{-1}(1)}, \dots, s_{\tilde{\mathcal{T}}_y^{-1}(N)}) \notag \\
 &=\mathrm{sign}(\tilde{\mathcal{T}}_y) \tilde{\psi}^{z_{\tilde{\mathcal{T}}_y(1)}, \dots, z_{\tilde{\mathcal{T}}_y(N)}}_0(s_1, \dots, s_N) \notag \\
 &=(-1)^{N_x} \delta_{\mathbf{s}} \chi_{\mathbf{s}} \prod_{i < j}^N \left(e^{2 \pi i / N_y} (z_i - z_j)\right)^{(s_i s_j + 1)/2} \notag \\
 &=(-1)^{N_x N/2} \tilde{\psi}^{z_1, \dots, z_N}_0(s_1, \dots, s_N).
\end{align}
Here, we have used that
\begin{align}
\label{eq:momentum-of-psi0-intermediate-result}
\prod_{i < j}^{N} e^{\frac{\pi i}{N_y} (s_i s_j + 1)} &= (-1)^{N_x \frac{N}{2} + N_x},
\end{align}
which follows from $\sum_{j=1}^N s_j = 0$.
The eigenvalue of $\psi_0$ with respect to $\mathcal{T}_y$ is therefore $(-1)^{N_x N/2}$.
With
\begin{align}
\label{eq:derivation-momentum-u}
\mathcal{T}_y u^a_{-n} \mathcal{T}_y^{-1} &= e^{- 2 \pi i n / N_y} u^a_{-n}
\end{align}
it follows that the eigenvalue of $\Jpsi$ is $e^{- 2 \pi i k / N_y} (-1)^{N_x N/2}$, where  $k = \sum_{j=1}^l n_j$.

For $\tilde{\psi}_0^{s_0, s_\infty, z_1, \dots, z_N}$, we obtain
\begin{align}
\label{eq:momentum-calculation-for-psi-0inf}
&\tilde{\psi}_0^{s_0, s_\infty, z_1, \dots, z_N}(s_{\tilde{\mathcal{T}}_y^{-1}(1)}, \dots, s_{\tilde{\mathcal{T}}_y^{-1}(N)}) \notag \\
&= (-1)^{N_x} \prod_{n=1}^{N} e^{\frac{\pi i}{N_y} (s_n s_0 + 1)} \prod_{n < m}^{N} e^{\frac{\pi i}{N_y} (s_n s_m + 1)} \notag \\
&\quad \times \tilde{\psi}_0^{s_0, s_\infty, z_1, \dots, z_N}(s_1, \dots, s_N) \notag \\
&= (-1)^{N_x + N_x \frac{N}{2}} \tilde{\psi}_0^{s_0, s_\infty, z_1, \dots, z_N}(s_1, \dots, s_N).
\end{align}
In the last equation, we have used that $\sum_{j=1}^N s_j + s_0 +
s_\infty = 0$.

\subsection{Inversion}
\label{sec:appendix-inversion-operator}
We require that the inversion $\mathcal{I}$
acts on the positions defined in Eq.~\eqref{eq:lattice-positions-on-complex-plane}
as
\begin{align}
\label{eq:action-I-on-z}
 z_{\tilde{\mathcal{I}}(i_x),\tilde{\mathcal{I}}(i_y)} = \frac{1}{z_{i_x,i_y}}.
\end{align}
This leads to the definition
\begin{align}
\label{eq:definition-of-inversion-as-permutation}
\tilde{\mathcal{I}}\left(i_x, i_y\right) &= \begin{cases} (N_x + 1 - i_x, N_y - i_y), &\text{if } i_y \neq N_y,\\
(N_x + 1 - i_x, N_y), &\text{if } i_y = N_y.
\end{cases}
\end{align}
We note that in our choice of $z_i$, the center of the cylinder is at
the unit circle. If this is not the case, then the definition of
Eq.~\eqref{eq:definition-of-inversion-as-permutation} leads to an
additional factor when $\tilde{\mathcal{I}}$ is applied to $z_i$.

In order to determine the sign of the permutation, we arrange the
state $| s_1, \dots, s_N \rangle$ in a matrix:
\begin{align}
\label{eq:spin-state-matrix}
      | s_{1, 1}, \dots, s_{N_x, N_y}\rangle &\cong
  \left(
  \begin{matrix}
  s_{1, 1} &\dots& s_{1, N_y}\\
  s_{2, 1} &\dots& s_{2, N_y}\\
  \vdots  &\vdots&  \vdots\\
  s_{N_x, 1}  &\dots&  s_{N_x, N_y}
  \end{matrix}
  \right).
\end{align}
The transformed state is then given by
\begin{align}
\label{eq:I-transformed-state}
&\mathcal{I} |s_{1,1}, \dots, s_{N_x, N_y} \rangle\notag \\&\cong
\left(
\begin{matrix}
s_{N_x, N_y - 1} & s_{N_x, N_y - 2} & \dots& s_{N_x, 1} & s_{N_x, N_y}\\
s_{N_x - 1, N_y - 1}&       s_{N_x - 1, N_y - 2} &\dots& s_{N_x - 1, 1} & s_{N_x - 1, N_y}\\
\vdots & \vdots  &\vdots&  \vdots &\vdots\\
s_{1, N_y - 1}  & s_{1, N_y - 2}  &\dots&  s_{1, 1} &s_{1, N_y}
\end{matrix}
\right).
\end{align}

To bring the transformed matrix back to the original form, we first
reverse all $N_y$ columns and then reverse all $N_x$ rows excluding
the last element of each row.
A single sequence of $L$ elements can be reversed in $\frac{1}{2} L (L-1)$ steps.
Therefore, the sign of the permutation is given by
\begin{align}
\label{eq:sign-of-I}
\mathrm{sign}(\tilde{\mathcal{I}}) &= (-1)^{N_y \frac{1}{2} N_x (N_x - 1) + N_x \frac{1}{2} (N_y - 1) (N_y - 2)}.
\end{align}

We next determine the contribution from the coordinate part of the wave function $\psi_0$.
Using Eq.~\eqref{eq:action-I-on-z}, we have
\begin{align}
\label{eq:transformation-of-coordinate-part}
&\tilde{\psi}^{z_{\tilde{\mathcal{I}}(1)}, \dots, z_{\tilde{\mathcal{I}}(N)}}_0(s_1, \dots, s_N) \notag \\
&= \tilde{\psi}^{z_{1}, \dots, z_{N}}_0(s_1, \dots, s_N) \prod_{m < n}^{N} (-z_m z_n)^{-\frac{1}{2}(s_m s_n + 1)} \notag \\
&= \tilde{\psi}^{z_{1}, \dots, z_{N}}_0(s_1, \dots, s_N) e^{-\frac{1}{4}  \sum_{m \neq n} (s_m s_n + 1) (\log(z_m z_n) + \pi i) } \notag \\
&= \tilde{\psi}^{z_{1}, \dots, z_{N}}_0(s_1, \dots, s_N) e^{-\frac{1}{4}  \sum_{m, n} (s_m s_n + 1) (\log(z_m z_n) + \pi i) } \notag\\
&\quad \times e^{\frac{1}{2}  \sum_{i} (2 \log(z_i) + \pi i) } \notag \\
&= \tilde{\psi}^{z_{1}, \dots, z_{N}}_0(s_1, \dots, s_N) (-1)^{\frac{N}{2} N_x + N_x}.
\end{align}
In the last step, we have used that $s_1 + \dots + s_N = 0$ and
\begin{align}
\label{eq:sum-log-positions}
\sum_{n=1}^{N} \log(z_n) &= \pi i \left(N + N_x\right).
\end{align}
Therefore, the eigenvalue of $\psi_0$ with respect to $\mathcal{I}$ is
\begin{align}
\label{eq:eigenvalue-psi0-with-resp-to-I}
\mathrm{sign}(\tilde{\mathcal{I}}) (-1)^{\frac{N}{2} N_x + N_x} &= (-1)^{\frac{N}{2} N_y}.
\end{align}

The states $\Jpsi$ are not eigenstates of $\mathcal{I}$, but
transform as
\begin{align}
\label{eq:transformation-of-Jpsi-with-respect-to-I}
\mathcal{I} | \Jpsi \rangle
&= \mathcal{I} u^{a_l}_{-n_l} \mathcal{I}^{-1} \dots  \mathcal{I} u^{a_1}_{-n_1} \mathcal{I}^{-1} \mathcal{I} | \psi_0 \rangle \notag\\
&= (-1)^{\frac{N}{2} N_y} u^{a_l}_{n_l} \dots u^{a_1}_{n_1} | \psi_0 \rangle.
\end{align}
Here, we have used that
\begin{align}
\label{eq:tranformation-of-single-u-under-I}
  \mathcal{I} u^{a_j}_{-n_j} \mathcal{I}^{-1} &
 = \sum_{i=1}^{N} \frac{1}{\left(z_i\right)^{n_j}} \mathcal{I} t^{a_j}_i \mathcal{I}^{-1}
 = \sum_{i=1}^{N} \frac{1}{\left(z_i\right)^{n_j}} t^{a_j}_{\tilde{\mathcal{I}}^{-1}(i)} \notag \\
 &= \sum_{i=1}^{N} \left(z_i\right)^{n_j} t^{a_j}_{i} = u^{a_j}_{n_j},
\end{align}
if the center of the cylinder is at the unit circle.
In terms of the states $\Jinfpsi$ defined in Eq.~\eqref{eq:definition-of-Jpsi-inf},
we therefore have
\begin{align}
  \label{eq:final-transformation}
  \mathcal{I} | \Jpsi \rangle = (-1)^{\frac{N}{2} N_y} | \Jinfpsi \rangle.
\end{align}
For the
transformed states $\mathcal{I} | \Jpsi \rangle$, the current operators are
therefore inserted at $z_\infty = \infty$ instead of at $z_0 = 0$.
Eigenstates of $\mathcal{I}$ with eigenvalues $(\pm 1) (-1)^{\frac{N}{2} N_y}$ are then
given by
\begin{align}
\label{eq:eigenstates-Jpsi-with-respect-to-I}
\Jpsi \pm \psi^{\phantom{-}a_l \dots \phantom{-}a_1}_{-n_l \dots -n_1}.
\end{align}

Finally, we determine the transformation of $\tilde{\psi}_0^{s_0, s_\infty, z_1, \dots, z_N}$ with respect to $\mathcal{I}$.
As for $\psi_0$, there is a contribution from the sign of the permutation
and from the transformation of the coordinates. The calculation is similar to that for $\psi_0$,
only that now $s_0 + s_\infty + \sum_{i=1}^{N} s_i = 0$. We find
\begin{align}
\label{eq:transformation-of-psi0inf-tilde-under-I}
  \mathcal{I} | \tilde{\psi}_0^{s_0, s_\infty, z_1, \dots, z_N} \rangle
  &= \mathrm{sign}(\tilde{\mathcal{I}}) | \tilde{\psi}_0^{s_0, s_\infty, z_{\tilde{\mathcal{I}}(1)}, \dots, z_{\tilde{\mathcal{I}}(N)}} \rangle \notag \\
  &= (-1)^{N_y \frac{N}{2} + N_x + 1} | \tilde{\psi}_0^{s_\infty, s_0, z_1, \dots, z_N} \rangle.
\end{align}
Since $\psi^{s_0, s_\infty}_0$ and $\tilde{\psi}_0^{s_0, s_\infty, z_1, \dots, z_N}$ only differ by a spin-independent factor,
we also have
\begin{align}
  \label{eq:transformation-of-psi0inf-under-I}
  \mathcal{I} | \psi_0^{s_0, s_\infty} \rangle &= (-1)^{N_y \frac{N}{2} + N_x + 1} | \psi_0^{s_\infty, s_0}\rangle.
\end{align}
Note that $\mathcal{I}$ exchanges the spins $s_0$ and $s_\infty$ in $\psi^{s_0, s_\infty}_0$.

\section{Exact parent Hamiltonians}
\label{sec:exact-parent-hamiltonians}
As shown in Sec.~\ref{sec:local-model-hamiltonians},
the edge states $\psi^a_1$ have a
good overlap with low-lying excited states of a local model, for which
$\psi_0$ approximates the ground state. In this section, we
analytically construct SU(2)-invariant, nonlocal parent Hamiltonians
for some linear combinations of the states $\Jpsi$,
i.e. Hamiltonians for which they are
exact eigenstates with the lowest energy.

\subsection{Construction of parent Hamiltonians}
The starting point of our construction is the operator
\begin{align}
  \label{eq:definition-of-C}
  \mathcal{C}^a &= \sum_{i \neq j}^{N}
                  \frac{z_i + z_j}{z_i - z_j} (t^a_j + i \vep_{abc} t^b_i t^c_j).
\end{align}
In Appendix~\ref{sec:appendix-action-of-c}, we explicitly compute the
action of $\mathcal{C}^a$ on states constructed from $\psi_0$ by
insertion of current operators, and show that $\mathcal{C}^a$ does not mix
the states $\Jpsi$ with different levels $k = \sum_{j=1}^l n_j$ if $k
< N_y$. This property is key to our construction of parent Hamiltonians:
It allows us to treat the levels separately starting with the
lower levels, which have fewer states. The action of $\mathcal{C}^a$
on states at level $k$ is described by a matrix. For low $k$,
the dimension of this matrix is considerably smaller compared
to that of an operator acting on the complete Hilbert space.
Moreover, the size of the matrix depends only on the level $k$
rather than the number of spins $N$.

We next add
a multiple of the total spin $T^a$ to $\mathcal{C}^a$ and define the
operators
\begin{align}
\label{eq:definition-of-D}
\mathcal{D}^a_n = \mathcal{C}^a + (n + 1 - N) T^a,
\end{align}
where $n$ is an integer. The operator $\mathcal{D}^a_n$ is also closed
in the subspace of states of level $k$ if $k < N_y$ since $T^a$ does
not mix states of different levels. For certain values of $n$,
we managed to find states constructed from current operators that are
annihilated by the three operators $\mathcal{D}^a_n$ for $a \in \{x,
y, z\}$.  These states are then ground states of the Hamiltonian
\begin{align}
\label{eq:definition-of-parent-Hamiltonian}
H_n &= \left(\mathcal{D}^a_n\right)^\dagger \mathcal{D}^a_n,
\end{align}
where the index $a$ is summed over. Note that the Hamiltonian $H_n$
is positive semi-definite and SU(2) invariant. $H_n$ is nonlocal and contains
terms with up to four-body interactions since $\mathcal{D}^a_n$ has terms
linear and quadratic in spin operators.

Before describing our results, we note that the condition $\mathcal{D}^a_n | \psi \rangle = 0$
for all $a$ implies that the state $\psi$ is part of the
subspace on which $T^b T^b$ and $\mathcal{D}^a_n$ commute.
To show this, we first note that
\begin{align}
\label{eq:commutator-D-and-T}
\left[\mathcal{D}^a_n, T^b\right] &= i \vep_{abc} \mathcal{D}^c_n,
\end{align}
which is a direct consequence of the definitions of Eqs.~\eqref{eq:definition-of-C}
and~\eqref{eq:definition-of-D}.
It then follows that
\begin{align}
\label{eq:subspace-commuting-tsq-and-d}
  \left[T^b T^b, \mathcal{D}^a_n\right] | \psi \rangle
  &= \left(-i \vep_{bac} T^b \mathcal{D}^c_n - i \vep_{bac} \mathcal{D}^c_n T^b\right) | \psi \rangle \notag\\
  & =-i \vep_{bac} \left[\mathcal{D}^c_n, T^b\right] | \psi \rangle
    = \vep_{bac} \vep_{cbd} \mathcal{D}^d_n | \psi \rangle \notag\\
  &= 0,
\end{align}
where we assumed that $\mathcal{D}^a_n | \psi \rangle = 0$ for all $a$.
The states satisfying $\mathcal{D}^a_n | \psi \rangle = 0$ can therefore be
decomposed into sectors of different total spin.

We note that the condition $\left[T^bT^b,
\mathcal{D}^a_n\right] | \psi \rangle = 0$ is equivalent to
\begin{align}
  \label{eq:tsq-d-commute-with-multiple-of-n}
  \left[T^bT^b, \mathcal{C}^a + (1-N) T^a\right] | \psi \rangle = 0,
\end{align}
since $T^bT^b$ and $T^a$ commute. The operator $\mathcal{C}^a + (1-N)
T^a$ has the advantage that its matrix entries in terms of the states
at level $k$ do not depend $N$ and $n$ [cf. Eq.~\eqref{eq:action-of-C}
in Appendix~\ref{sec:appendix-action-of-c}\,]. In our calculations, we
found it technically easier to first determine the subspace of states
on which $T^b T^b$ and $\mathcal{C}^a + (1 - N) T^a$ commute and then look for
states that are annihilated by $\mathcal{D}^a_n$ for a suitable $n$
within that subspace.

\begin{table}[htb]
  \caption{
    \label{tab:ground-states-analytical}States constructed from
current operators that are annihilated by $\mathcal{D}^a_n$ for $a \in
\{x, y, z\}$ on a cylinder with $N_y > k$
[cf. Eqs.~\eqref{eq:definition-of-C} and \eqref{eq:definition-of-D}
for the definition of $\mathcal{D}^a_n$].  For $N_y$ sufficiently
large ($N_y > k$), these states are ground states of the Hamiltonian
$H_n = \left(\mathcal{D}^a_n\right)^\dagger \mathcal{D}^a_n$.}
\centering
\begin{ruledtabular}

\begin{tabular}{llll}
 $k$ &                                                                                                                                                                                                                                                      State & Spin &  $n$ \\
\hline
 $0$ &                                                                                                                                                                                                                                                   $\psi_0$ &  $0$ &  any \\
 $1$ &                                                                                                                                                                                                                                             $\psi^{a}_{1}$ &  $1$ &  $1$ \\
 $2$ &                                                                                                                                                                                                                                                        --- &  --- &  --- \\
 $3$ &                                                                                                                                                                                         $\psi^{a}_{3} + i \varepsilon_{a d e}  \psi^{d\phantom{,}e}_{2,1}$ &  $1$ &  $5$ \\
 $4$ &                                                                                                                                                                                                   Symmetric-traceless part of $\psi^{a\phantom{,}b}_{3,1}$ &  $2$ &  $3$ \\
 $5$ &                                                      $\psi^{a\phantom{,}d\phantom{,}d}_{3,1,1} + \frac{3}{2} i \varepsilon_{a d e}  \psi^{d\phantom{,}e}_{3,2} + \frac{3}{2} i \varepsilon_{a d e}  \psi^{d\phantom{,}e}_{4,1} + \frac{9}{4} \psi^{a}_{5}$ &  $1$ &  $9$ \\
 $6$ &                                                                                                                                                                                                                                                        --- &  --- &  --- \\
 $7$ & $\psi^{a\phantom{,}d\phantom{,}d}_{3,3,1} + 4 \psi^{d\phantom{,}a\phantom{,}d}_{4,2,1} + \frac{5}{3} \psi^{a\phantom{,}d\phantom{,}d}_{4,2,1} + \frac{7}{3} \psi^{a\phantom{,}d\phantom{,}d}_{5,1,1} +  i \varepsilon_{a d e}  \psi^{d\phantom{,}e}_{4,3}$ &  $1$ & $13$ \\
     &                                                                                              $+ \frac{5}{2} i \varepsilon_{a d e}  \psi^{d\phantom{,}e}_{5,2} + \frac{9}{2} i \varepsilon_{a d e}  \psi^{d\phantom{,}e}_{6,1} + \frac{21}{4} \psi^{a}_{7}$ &      &      \\
 $8$ &                                                                                                                                                                                                                                Symmetric-traceless part of &  $2$ &  $7$ \\
     &          $i \varepsilon_{a d e} \psi^{b\phantom{,}d\phantom{,}e}_{4,3,1} + \frac{1}{2} i \varepsilon_{a d e} \psi^{b\phantom{,}d\phantom{,}e}_{5,2,1} - \frac{1}{2} \psi^{a\phantom{,}b}_{5,3} - \psi^{a\phantom{,}b}_{6,2} -2 \psi^{a\phantom{,}b}_{7,1}$ &      &      \\
 $9$ &                                                                                                                                                                                     Symmetric-traceless part of $\psi^{a\phantom{,}b\phantom{,}c}_{5,3,1}$ &  $3$ &  $5$ \\
 $9$ &        $i \varepsilon_{a d e} \psi^{d\phantom{,}e\phantom{,}f\phantom{,}f}_{4,3,1,1} - \frac{1}{2} \psi^{a\phantom{,}d\phantom{,}d}_{4,3,2} - \frac{3}{2} \psi^{a\phantom{,}d\phantom{,}d}_{5,2,2} - \frac{1}{2} \psi^{a\phantom{,}d\phantom{,}d}_{4,4,1}$ &  $1$ & $17$ \\
     &                                        $+ 3 \psi^{d\phantom{,}d\phantom{,}a}_{5,3,1} - \frac{9}{2} \psi^{d\phantom{,}a\phantom{,}d}_{5,3,1} - \frac{9}{2} \psi^{a\phantom{,}d\phantom{,}d}_{5,3,1} - \frac{9}{2} \psi^{d\phantom{,}a\phantom{,}d}_{6,2,1}$ &      &      \\
     &                     $- \frac{7}{2} \psi^{a\phantom{,}d\phantom{,}d}_{6,2,1} -3 \psi^{a\phantom{,}d\phantom{,}d}_{7,1,1} - \frac{27}{8} i \varepsilon_{a d e}  \psi^{d\phantom{,}e}_{5,4} - \frac{21}{8} i \varepsilon_{a d e}  \psi^{d\phantom{,}e}_{6,3}$ &      &      \\
     &                                                                                           $- \frac{45}{8} i \varepsilon_{a d e}  \psi^{d\phantom{,}e}_{7,2} - \frac{63}{8} i \varepsilon_{a d e}  \psi^{d\phantom{,}e}_{8,1} - \frac{105}{8} \psi^{a}_{9}$ &      &      
\end{tabular}
\end{ruledtabular}
\end{table}

We summarize our analytical results in
Table~\ref{tab:ground-states-analytical}. The states with spin $2$ and
$3$ appear as the symmetric-traceless parts of states with $2$ and $3$
open indices, respectively. For a two-index state $\phi^{ab}$, the
symmetric-traceless part is defined as
\begin{align}
  \label{eq:symmetric-traceless-two-open}
  3 (\phi^{ab} +  \phi^{ba}) - 2 \delta_{ab} \phi^{dd},
\end{align}
and for a three-index state $\phi^{abc}$ as\cite{Spencer1970}
\begin{align}
  \label{eq:symmetric-traceless-three-open}
  &5 \big(\phi^{abc} + \phi^{bca} + \phi^{cab} + \phi^{cba} + \phi^{bac} + \phi^{acb}\big) \notag \\
  -&2 \big(\delta_{ab} (\phi^{cdd} + \phi^{dcd} + \phi^{ddc}) + \delta_{ac} (\phi^{bdd} + \phi^{dbd} + \phi^{ddb}) \notag \\
  &\phantom{2\big(}+ \delta_{bc} (\phi^{add} + \phi^{dad} + \phi^{dda})\big).
\end{align}
Except for the levels $2$ and $6$, we find states and corresponding
parent Hamiltonians for all levels that were considered.
Note that the singlet
$\psi_0$ is a ground state of $H_n$ for any value of $n$.  For the
additional ground states, we observe that the value of $n$ tends to be
larger at higher levels $k$.
This means that the ground state space of the Hamiltonians $H_n$
with lower $n$ contains states of a lower level in current
operators.  For example, we only find the ground states
$\psi_0$ and $\psi^a_{1}$ for $H_1$.  Similarly, the only appearing ground
states of $H_3$ at levels $k \le 9$ are $\psi_0$ and the
symmetric-traceless part of $\psi^{a\phantom{,}b}_{3, 1}$.

\subsection{Ground-state degeneracies}
In the previous subsection, we explicitly constructed
analytical ground states of the Hamiltonians $H_n$ with $n \in \{1,
   3, 5, 7, 9, 13, 17\}$ in terms of linear combinations of states
$\Jpsi$ with levels $k \le 9$.  We now study the ground state
spaces of the Hamiltonians $H_n$ numerically and provide evidence for
$n \in \{1, 3, 5\}$ that the complete ground state space is spanned
by the states given in Table~\ref{tab:ground-states-analytical}.

\begin{table}[htb]
\caption{\label{tab:ground-states-numerical}Numerically determined
ground state multiplets of the Hamiltonians $H_n$ for $n \le 13$ and
an even number of spins $N$ with $N \le 14$. The second column
indicates the minimal number of spins $\NyMin$ in the periodical
direction for which the complete shown multiplet was observed in all
system with $\NyMin \le N_y \le 14$.  For a lower number of spins in the
$y$-direction, the observed ground state space is smaller. For even
$n$, we only find a singlet ground state.}
\centering
\begin{ruledtabular}

\begin{tabular}{lll}
  $n$ & $\NyMin$ &                  Ground state multiplet \\
\hline
  $1$ &      $2$ &                            $0 \oplus 1$ \\
  $3$ &      $4$ &                            $0 \oplus 2$ \\
  $5$ &      $6$ &                   $0 \oplus 1 \oplus 3$ \\
  $7$ &      $8$ &                   $0 \oplus 2 \oplus 4$ \\
  $9$ &     $10$ &          $0 \oplus 1 \oplus 3 \oplus 5$ \\
 $11$ &     $12$ &          $0 \oplus 2 \oplus 4 \oplus 6$ \\
 $13$ &     $14$ & $0 \oplus 1 \oplus 3 \oplus 5 \oplus 7$ 
\end{tabular}
\end{ruledtabular}
\end{table}

By an exact diagonalization, we numerically determined the
ground state multiplets of the Hamiltonians $H_n$ for $n \le 13$ and
systems with $N = N_x N_y \le 14$ and $N$ even.
Our results are summarized in Table~\ref{tab:ground-states-numerical}.
We observe that states with spin $s$ occur in the ground state spaces only in
systems with $N_y \ge 2 s$. Furthermore, we find that the ground state
degeneracy does not grow anymore if $N_y$ reaches a certain value $\NyMin$.
This statement is most conclusive for the lower values $n$, where $\NyMin$
is smaller and we are thus able to probe more systems with $N_y \ge \NyMin$.
For $n \in \{1, 3, 5\}$, this implies that all ground states are given
by the corresponding states of Table~\ref{tab:ground-states-analytical}.

Finally, let us formulate a conjecture about the structure of
the states annihilated by $\mathcal{D}^a_n$, which are
ground states of $H_n$.
Our analytical results are consistent with the following rule: For each
spin sector $s \in \{1, 2, \dots \}$, there is a series of states at levels $k = s^2 + 2 s j$
with $j \in \{0, 1, 2, \dots\}$.
These states are annihilated by $\mathcal{D}^a_{n}$ with $n = 2 s - 1 + 4 j$.
As one can show by induction, the second rule implies that the ground state
space of $H_n$ with $n = 2 s - 1$ contains the multiplet
\begin{align}
0 \oplus \begin{cases}
1 \oplus 3 \oplus \dots \oplus s, &\text{if } s \text{ is odd}, \notag \\
2 \oplus 4 \oplus \dots \oplus s, &\text{if } s \text{ is even.}
\end{cases}
\end{align}
The numerical results of Table~\ref{tab:ground-states-numerical}
are consistent with this multiplet structure and thus support the conjecture
that the values of $n$ are given by $n = 2 s - 1 + 4 j$.

\section{Action of $\mathcal{C}^a$ on states built from current operators}
\label{sec:appendix-action-of-c}
Our starting point is the decoupling equation for
the states $\JpsiOne$ derived in Ref.~\onlinecite{Herwerth2015}. This
equation describes the action of the operator
\begin{align}
\label{eq:operator-ci-definition}
\mathcal{C}^a_i = \sum_{j \in \{1, \dots, N\} \setminus \{i\}} \frac{z_i + z_j}{z_i - z_j} (t^a_j + i \vep_{abc} t^b_i t^c_j)
\end{align}
on the states $\JpsiOne$ and follows from the CFT null field
\begin{align}
\label{eq:null-field-and-definition-of-k}
(K^a_b)_i (J^b_{-1} \phi_{s_i})(z_i) &\quad\text{with}\quad (K^a_b)_i = \delta_{ab} - i \vep_{abc} t^c_i.
\end{align}
[The definition of $\mathcal{C}^a_i$ used here differs from that
of Ref.~\onlinecite{Herwerth2015} by a factor of $2/3$.]
\begin{widetext}
The operator
\begin{align}
\label{eq:operator-c-definition}
\mathcal{C}^a &= \sum_{i=1}^{N} \mathcal{C}^a_i
\end{align}
was used in Sec.~\ref{sec:exact-parent-hamiltonians} to construct parent Hamiltonians
for states built from current operators.

The decoupling equation reads
\begin{align}
\label{eq:decoupling-equation-Jpsi}
\mathcal{C}_i^a | \psi^{a_k\dots a_1}_{1\phantom{_k} \dots 1} \rangle
  &= \sum_{q=1}^k\frac{(K_{a_q}^a)_i}{z_i} | \psi^{a_k\dots a_{q+1} a_{q-1}\dots a_1}_{1\phantom{_k} \dots 1\phantom{_{q + 1}} 1\phantom{_{q-1}} \dots 1} \rangle
  +(K_{b}^a)_i T^b | \psi^{a_k\dots a_1}_{1\phantom{_k} \dots 1} \rangle \notag \\
  &\quad+2 (K_{b}^a)_i \sum_{s_1, \dots, s_N}
    \sum_{q=2}^k\sum_{n=0}^{q-1} \frac{i\vep_{ba_qc}}{z_i^{n+1}} \langle \PhiString (J_{-1}^{a_k}\dots J_{-1}^{a_{q+1}} J_n^{c}J_{-1}^{a_{q-1}} \dots J_{-1}^{a_1})(0)\rangle
    | s_1, \dots, s_N \rangle,
\end{align}
where
\begin{align}
\label{eq:definition-of-phistring}
\PhiString &= \phi_{s_1}(z_1) \dots \phi_{s_N}(z_N).
\end{align}
The decoupling equation for the states $\JpsiOne$, where all mode numbers
are one, is enough to describe
the action of $\mathcal{C}^a_i$ on states with general mode numbers $\Jpsi$: Using the
Kac-Moody algebra of Eq.~\eqref{eq:Kac-Moody-current-algebra},
the latter can be expressed in terms of the states $\JpsiOne$
by repeated application of
\begin{align}
\label{eq:elim-higher-order-J}
J^a_{-n} &= \frac{i}{2} \vep_{abc} \left[J^{c}_{-1}, J^{b}_{-n+1}\right] \quad(n \neq 0).
\end{align}
On the cylinder, we have
\begin{align}
\label{eq:summation-over-powers-of-z}
\sum_{i=1}^{N} (z_i)^{-n} &= 0,  \quad\text{if } n \text{ mod } N_y \neq 0.
\end{align}
Summing over $i$ in Eq.~\eqref{eq:decoupling-equation-Jpsi}, we therefore obtain for $k < N_y$
\begin{align}
\label{eq:action-of-C}
  C^a | \JpsiOne \rangle &= (N-1) T^a | \JpsiOne \rangle + \sum_{q=1}^k i \vep_{a_q a c}
| \psi^{c a_{k} \dots a_{q+1} a_{q-1} \dots a_{1}}_{1 1\phantom{_{k}} \dots 1\phantom{_{q+1}} 1\phantom{_{q-1}} \dots 1} \rangle \notag \\
 & \quad + \sum_{s_1, \dots, s_N} \sum_{q=2}^k \sum_{n=0}^{q-1} G^{q, n}_{a_k \dots a_1}(s_1, \dots, s_N) | s_1, \dots, s_N \rangle,
\end{align}
with
\begin{align}
\label{eq:G-symbol-definition}
G^{q, n}_{a_k \dots a_1}(s_1, \dots, s_N) &= 2 \langle \PhiString (J^{a_q}_{-n-1} \Ja{k}
  \dots \Ja{q+1} J^a_n \Ja{q-1} \dots \Ja{1})(0) \rangle \notag \\
  & \quad - 2 \delta_{a_q
a} \langle \PhiString (J^c_{-n-1} \Ja{k} \dots \Ja{q+1} J^{c}_n
\Ja{q-1} \dots \Ja{1})(0) \rangle.
\end{align}
The first two terms on the right hand side of Eq.~\eqref{eq:action-of-C} are of order $k$ in current operators since
\begin{align}
  \label{eq:total-spin-on-Jstate}
  T^a | \JpsiOne \rangle = \sum_{q=1}^k i \vep_{a a_q c} | \psi^{a_{k} \dots a_{q+1} c a_{q-1} \dots a_{1}}_{1\phantom{_{k}} \dots 1\phantom{_{q+1}} 1 1\phantom{_{q-1}} \dots 1} \rangle.
\end{align}
In the remaining terms, the modes $J^a_n$ and $J^c_n$
can be commuted to the right since $J^a_n | 0 \rangle = 0$ for $n \ge 0$:
\begin{align}
  \label{eq:commute-Jan-to-the-right}
  (J^a_n \Ja{q-1} \dots \Ja{1})(0) | 0 \rangle = \sum_{r=1}^{q-1} i \vep_{a a_r d} (\Ja{q-1} \dots \Ja{r+1} J^d_{n-1} \Ja{r-1} \dots \Ja{1})(0) | 0 \rangle,
\end{align}
and similarly for $(J^{c}_n \Ja{q-1} \dots \Ja{1})(0) |0
\rangle$. Iterating this step, the current operator modes with a
positive mode number can be eliminated. The resulting terms only have
negative mode numbers and are all of order $k$ in current operators.
\end{widetext}

\bibliography{refs}

\end{document}